\documentclass[a4paper,fleqn,usenatbib]{mnras}

\usepackage[T1]{fontenc}
\usepackage{ae,aecompl}

\usepackage{graphicx}	
\usepackage{amsmath}	
\usepackage{amssymb}	
\usepackage{multicol}
\usepackage{xspace}
\usepackage{xcolor}
\usepackage[normalem]{ulem}

\makeatletter

\newcommand{\Rmnum}[1]{\expandafter\@slowromancap\romannumeral #1@}
\makeatother

\newcommand{\sage}{{\scriptsize SAGE}\xspace} 
\newcommand{\hen}{{\scriptsize L-GALAXIES}\xspace} 
\newcommand{\fpass}{$\rm f_{passive}$\xspace} 

\newcommand{\Sec}[1]{{\protect\hyperref[sec:#1]{Section~\ref*{sec:#1}}}}
\newcommand{\App}[1]{{\protect\hyperref[sec:#1]{Appendix~\ref*{sec:#1}}}}
\newcommand{\Fig}[1]{{\protect\hyperref[fig:#1]{Fig.~\ref*{fig:#1}}}}
\newcommand{\Equ}[1]{{\protect\hyperref[equ:#1]{Eq.~\ref*{equ:#1}}}}
\newcommand{\Tab}[1]{{\protect\hyperref[tab:#1]{Table~\ref*{tab:#1}}}}

\title[Quenching in Local Central Galaxies]{On The Role of Supermassive Black Holes in Quenching Star Formation in Local Central Galaxies}

\author[N. Arora et al.]{
Nikhil Arora,$^{1}$\thanks{E-mail: nikhil.arora@queensu.ca}
Matteo Fossati,$^{2}$
Fabio Fontanot,$^{3}$
Michaela Hirschmann,$^{4}$
\newauthor
David J. Wilman,$^{5,6}$
\\
$^{1}$Department of Physics, Engineering Physics and Astronomy, Queen's University, Kingston, ON K7L 3N6, Canada\\
$^{2}$Institute for Computational Cosmology and Centre for Extragalactic Astronomy, Department of Physics, Durham University, \\
South Road, Durham DH1 3LE, UK\\
$^{3}$INAF - Astronomical Observatory of Trieste, via G.B. Tiepolo 11, I-34143 Trieste, Italy\\
$^{4}$DARK, Niels Bohr Institute, University of Copenhagen, Lyngbyvej 2, DK-2100 Copenhagen, Denmark\\
$^{5}$Universit\"ats-Sternwarte M\"unchen, Scheinerstrasse 1, D-81679 M\"unchen, Germany\\
$^{6}$Max-Planck-Institut f\"ur Extraterrestriche Physik, Giessenbachstrasse, D-85748 Garching, Germany\\
}

\date{Accepted XXX. Received YYY; in original form ZZZ}

\pubyear{2019}

\begin{document}
\label{firstpage}
\pagerange{\pageref{firstpage}--\pageref{lastpage}}
\maketitle


\defcitealias{henriques14}{H15}
\defcitealias{croton16}{C16}


\begin{abstract}
  In this work, we analyze the role of AGN feedback in quenching star formation for massive, central galaxies in the local Universe.
  In particular, we compare the prediction of two semi-analytic models (\hen and \sage) featuring different schemes for AGN feedback, with the SDSS DR7 taking advantage of a novel technique for identifying central galaxies in an observational dataset.
  This enables us to study the correlation between the model passive fractions, which is predicted to be suppressed by feedback from an AGN, and the observed passive fractions in an observationally motivated parameter space.
  While the passive fractions for observed central galaxies show a good correlation with stellar mass and bulge mass, passive fractions in \hen correlate with the halo and black hole mass.
  For \sage, the passive fraction correlate with the bulge mass as well. Among the two models, SAGE has a smaller scatter in the black hole - bulge mass $(\rm {M_{BH}-M_{Bulge}})$ relation and a slope that agrees better with the most recent observations at $z \sim 0$.
  Despite the more realistic prescription of radio mode feedback in \sage, there are still tensions left with the observed passive fractions and the distribution of quenched galaxies.
  These tensions may be due to the treatment of galaxies living in non-resolved substructures and the resulting higher merger rates that could bring cold gas which is available for star formation.

\end{abstract}





\section{Introduction}
\label{sec:intro}

The interplay between gravitational and hydrodynamic processes dictate the formation and evolution of galaxies in the Universe.
Large amounts of dark matter, through gravitational interactions, form halos that provide a gravitational potential for the baryons (gas and stars) to fall towards their cores.
This gas, through its self-gravity and radiative processes, cools down to form compact clouds that lead to the formation of stars, forming extended structures called galaxies that evolve in time through gravitational and hydrodynamic interactions \citep{white91}.
It is now widely accepted that the formation of galaxies and clusters of galaxies occurs through hierarchical clustering of matter in the $\Lambda$ Cold Dark Matter ($\Lambda$CDM) paradigm \citep{white&rees, white91}.
With respect to the hydrodynamic processes, galaxies can be broadly split into two categories; star forming galaxies which appear blue in the sky and quiescent galaxies that are red and do not form stars at present times \citep{dressler1980, baldry2006}.
This bimodality in population is evident in a relation that connects the Star Formation Rate (SFR) and the stellar mass of the galaxy \citep{whitaker2012, cano2016, santini}.
The star forming galaxies lie on a tight relation whereas the quiescent galaxies form a population below the relation. 

``Quenching", i.e. the combination of physical and dynamical processes leading to the fast decrease of star formation activity in a galaxy and its removal from the SFR-stellar mass relation, as a function of various galaxy structural and dynamical parameters have been widely studied over the past few decades.
Using the Sloan Digital Sky Survey (SDSS) data, it has been demonstrated that the galaxy colour bimodality strongly depends on stellar mass and the environment \citep{baldry2006, wilman10, peng2010, peng2012}.
In particular \cite{bluck14}, show that fraction of passive galaxies (\fpass) as a function of stellar mass with the bulge mass.
Using the tight relation between the bulge mass and Supermassive Black Hole (SMBH) mass reported in \cite{haring04}, \cite{bluck2014, bluck14} argue that \fpass should on the black hole mass and hence the Active Galactic Nuclei (AGN) luminosity.
\cite{withaker}, with 3D-HST data, studied the relationship between the SFR, galaxy sizes and central densities for high redshift galaxies.
They find that galaxies with high central densities are red and have lower specific star formation rate $\rm sSFR (=SFR/M_{*})$ whereas galaxies with low surface central density are blue and have, on average, higher $\rm sSFR$.

In theoretical models of galaxy formation within a $\Lambda$CDM Universe, massive galaxies reside the centre of galaxy groups and clusters.
At late times, such systems are supposed to grow in mass through accretion from cooling flows which fuels star formation.
These cooling flows would lead to continuous growth, making central galaxies more massive and compact (high s\'ersic indices) \citep{croton06, Donzelli2011, Cooper2015}.
This is in contrast with the observed galaxy stellar mass function (SMF) and luminosity function which depicts a knee at the high mass end \citep{benson03, daalen17}.
This cut-off at high mass suggests the presence of a mechanism that either removes the gas or prevents it from cooling down, making galaxies red and dead.
A number of physical mechanisms have been proposed to explain how quenching of star formation is ensued and sustained in galaxies.
At their core, these mechanism involve either heating, ionizing or stripping the gas from the galaxy \citep{gabor2010}.
For massive galaxies, called central galaxies, the supermassive black hole (SMBH) plays a critical role in regulating/halting star formation.
The energy created by the SMBH, referred to as Active Galactic Nuclei (AGN) feedback, has the potential to heat, ionize or eject the cold gas from the galaxy \citep{somerville2015, king15}.

The energy and momentum output from a SMBH, called AGN feedback, can affect the gas in three ways; heating the gas (thermal feedback); ionizing or photo-dissociate the gas (radiative feedback); or ejecting the gas through the presence of hot gas bubbles, winds or jets (mechanical feedback) \citep{somerville2015}.
\cite{DiMatteo2005} and \cite{springel05} carried out 3D simulations of AGN feedback and showed that depositing $\sim 5\%$ of the AGN bolometric luminosity in the surrounding gas particles can lead of very strong galactic outflows that halt the black hole growth and remove almost all gas from the galaxy, quenching star formation.
These simulations lacked cosmological initial conditions and consider the sole case of a binary galactic merger of ideal disk galaxies with no hot gas halos.
However, these simulations still produce self-regulated BH growth and tight $M_{BH}-\sigma$ relation, which matches observations \citep{tremaine2002, allesandra2012}.

Semi analytic models (SAMs) have also developed schemes to apply separate prescriptions of radiative feedback through winds and radio mode feedback through jets.
In a landmark study, \cite{croton06} introduced two modes of AGN feedback: quasar mode where the accretion is comparable to the Eddington limit and radio mode with radiatively inefficient accretion.
The energy output from radio mode feedback is then used to regulate BH growth and create a hot gas halo.
Recipes for AGN feedback differ from model to model but \cite{somerville2015} identified some common features in AGN feedback schemes for different models.
All BHs grow through cooling flows that results in accretion of hot and cold gas.
BH accretion is simulated through instabilities created in the disk or due to mergers, where accretion is radiatively efficient.
Radiatively inefficient accretion leads to low energy jets.
The energy output by these low energy jets are proportional to the mass of the BH and is used to offset cooling flows and govern heating of the gas \citep{croton06, somerville08, fontanot06, fontanot11, guo, hirschmann14, hirschmann16}.
The heating and cooling processes of the gas in these SAMs are calculated independently and hence are decoupled.
\cite{croton16} proposed a coarse way to couple the above mentioned heating and cooling processes.
They assume that cold gas is heated by the AGN within a radius $r_{heat}$ which is proportional to the heating and cooling rates.
The gas in that region never cools again.

Observational evidence for AGN feedback is still very weak.
Brightest Cluster Galaxies (BCG) offer the best evidence for the presence of AGN.
Without feedback BCGs would go through more star formation events \citep{fabian2012}.
X-ray observations of central cluster galaxies point to the presence of hot gas atmospheres that have very large cooling times which are associated with mechanical feedback from AGN activity\citep{fabian94, hogan17}.

It is also possible to quench massive isolated galaxies due starvation of gas.
Galaxies with $\rm{M_{*}>10^{11}\, M_{\odot}}$ seem to have very low fraction of neutral hydrogen, $\rm{f_{HI}\equiv \log_{10}(M_{HI}/M_{*}) < -2}$ \citep{huang2012}.
These galaxies are expected to have low star formation activity.
Furthermore, galaxies with massive bulges have disks which have a high Toomre Q parameter which prevents neutral gas collapse leading to morphological quenching \citep{kennicutt1989, martig2009}.
\cite{b2019} used the CALIFA \citep{califa} galaxies to show that systems with bulge-to-total luminosity ratios greater than $0.2$ are predominantly found to be quenched.

The aim of this paper is to study how SMBH processes control quenching of star formation in central galaxies in the local universe.
For such a task, a pure and comparable selection of massive central galaxies from SAMs and observational data is of the utmost importance.
To uniformly select central galaxies from both SAMs and observational data, we apply a modified techniques developed in \cite{fossati2015} of assigning stellar mass ranks within a cylindrical aperture.
For these centrals, we calculate the fraction of quenched galaxies within a given bin.
It has been reported that such quenched fractions depend very strongly on the stellar mass ($\rm M_{stellar}$) and the dark matter halo mass ($\rm M_{halo}$).
However, due to the strong correlation between the two quantities \citep{matthee, gu2016} and the complexities involved in calculating $\rm M_{halo}$ for observational galaxies, a new, more observationally motivated parameter is needed.
The projected density (number of neighbours in a cylindrical aperture) can be calibrated easily and correlates strongly with the $\rm M_{halo}$ at fixed stellar mass \citep{hogg03, kaufmann04, croton05, wilman10}.
Therefore, we will also test the trends of the passive fractions for central galaxies in a more observationally motivated parameter space which includes stellar mass and density of neighbours in a cylindrical aperture.

This paper is structured in the following way.
In Section 2, we describe the galaxy formation models we use accompanied with a detailed description of cooling modes and AGN feedback prescriptions which is followed by describing the observational data, the SDSS(DR7), section 3.
In Section 4, we discuss the parameter space that can uniformly be used between observations and SAMs to study star formation quenching.
In Section 5, we describe the algorithm implement to select massive central galaxies from both model and observed galaxies.
Section 6 presents passive fraction in central galaxies and it's correlation with various halo driven and baryonic driven galaxy properties.
In Section 7, we discuss the black hole - bulge mass relation and star formation quenching in that parameter space.
Finally, section 8 presents the conclusions and the global interpretation of our results.


\section{Galaxy Formation Models}

For this study, we use the SAMs provided in \cite[][hereafter H15]{henriques14} and \cite[][hereafter C16]{croton16}.
Both SAMs adopt an AGN feedback prescriptions that correspond to an improved version of \cite{croton06} model.
In this section, we briefly describe the updates to the galaxy formation models and the radio mode AGN feedback prescriptions.


\subsection{L-GALAXIES}

We start with the version of Munich SAM described in \citetalias{henriques14} which is an update of the model of \cite{guo}.
The galaxy formation models has been implemented on Millennium \citep{Springel2005} and Millennium-\Rmnum{2} \citep{bk2009} dark matter simulation to achieve a range of five order of magnitudes in stellar masses $(10^{7}M_{\odot}<M_{*}<10^{12}M_{\odot})$.
Model galaxies output match the abundance of galaxies and their passive fraction from $z=3$ to $z=0$.
\citetalias{henriques14} adopt the \cite{planck14} cosmology; $\sigma_8=0.829$, $H_0=67.3\rm{km~s^{-1}}$, $\Omega_{\Lambda}=0.685$, $\Omega_{m}=0.315$, $\Omega_{b}=0.0487~(f_b=0.155)$ and $n=0.96$.
The Munich SAM has updated treatment of the baryonic processes to address two problems; (1) very early formation and quenching of low-mass galaxies, and (2) large fraction of massive galaxies still forming stars at low redshift.
These problems are solved by delaying the reincorporation of the wind ejecta, lowering the threshold surface density of cold gas for star formation, eliminating ram pressure stripping in halos with mass less than $M\sim 10^{14}M_{\odot}$.
Furthermore,\citetalias{henriques14} uses the radio mode AGN feedback scheme from \cite{croton06} to efficiently suppress gas cooling and star formation at lower redshift.

\subsubsection{Radio mode feedback}

Previous versions of Munich SAMs used the same radio mode feedback prescription as in \cite{croton06}.
However, the feedback model still results in a large amount of massive star forming systems at $z=0$ \citep{henriques2013}.
In \citetalias{henriques14}, the radio mode feedback is modified to suppress cooling and star formation more efficiently at late times.
The continual accretion of hot gas from the host galaxies is formulated to be

\begin{equation}
\rm{\dot{M}_{BH, R}=k_{AGN}\Bigg(\dfrac{M_{hot}}{10^{11}M_{\odot}}\Bigg)\Bigg(\dfrac{M_{BH}}{10^{8}M_{\odot}}\Bigg).}
\label{equ:radiomode}
\end{equation}

In \Equ{radiomode}, $\rm M_{hot}$ is the hot gas mass and $\rm M_{BH}$ is the mass of the black hole in the host galaxy, $\rm k_{AGN}$ is the normalization of the radio mode feedback with a value of $\rm{5.3\times 10^{-3}~M_{\odot}yr^{-1}}$ (see \citetalias{henriques14}).

The accretion of material onto the SMBH results in energy injected into the halo in the form of jets. The energy in the jets is 

\begin{equation}
\rm{\dot{E}_{radio}=\eta\dot{M}_{BH, R}c^2,}
\label{equ:lum_BH}
\end{equation}

where $\rm{\eta=0.1}$ is efficiency parameter and $\rm c$ is the speed of light.
The energy from the jet modifies the cooling rate of the gas disk by

\begin{equation}
\rm{\dot{M}_{\rm{cool, eff} }= \max [\dot{M}_{cool} - 2\dot{E}_{radio}/V^2_{200c}, 0].}
\label{equ:cooling_rate}
\end{equation}

These jets add hot gas to the surrounding to suppress cooling and therefore star formation.
These massive systems use up all the available cold gas and then can no longer acrete cold gas leading to quenching.


\subsection{SAGE}
\label{sec:sage}

\citetalias{croton16} presented the Semi-Analytic Galaxy Evolution (or \sage) model which is an update of the SAM presented in \cite{croton06}.
The galaxy formation model updates a number of physical prescriptions: gas accretion, ejection due to feedback, reincorporation via the galactic fountain, gas cooling-radio mode AGN heating cycle, quasar mode AGN feedback, treatment of gas in satellite galaxies, galaxy merger and disruption and build-up of intra-cluster stars.
For this study, we use a galaxy catalogue from the Theoretical Astrophysical Observatory\citep{bernyk2016}\footnote{\url{https://tao.asvo.org.au/tao/}} where \sage is applied to the Millennium simulation \citep{Springel2005} with WMAP-1 cosmology \citep{wmap1};  $\sigma_8=0.9$, $H_0=73.0\rm{km~s^{-1}}$, $\Omega_{\Lambda}=0.75$, $\Omega_{m}=0.25$, $\Omega_{b}=0.045~(f_b=0.17)$ and $n=1.0$.
The fiducial parameters are constrained primarily to the stellar mass function at $z=0$ from \cite{baldry2008}.
Furthermore, a secondary set of constrains are applied using the star formation rate density history \citep{somerville2001}, the baryonic Tully-Fisher relation \citep{stark2009}, the mass-metallicity relation \citep{tremonti2004} and the black hole-bulge mass relation \citep{scott2013}.

Another important feature to note in \sage is the treatment of galaxies whose parent dark matter substructures are lost below the mass resolution limit of the Millennium Simulation (i.e. the so-called orphan galaxies).
\sage doesn't follow the evolution of such a population, but it is assumed that these objects are instantaneously disrupted due tidal interactions.
The stellar mass from the disrupted galaxy get added to either the intra-cluster component or gets added to the central galaxy.
This decision depends on duration of survival of the subhalo with respect to the average for a subhalo of its general properties.
Such a treatment can result in substantially altering the stellar masses of central galaxies and consequently the shape of the SMF (\citetalias{croton16}, \cite{knebe18}).
The reader is referred to \citetalias{croton16} for further details.

\subsubsection{Radio mode feedback}
Radio mode feedback prescription in \sage is an update from the model presented in \cite{croton06}.
The accretion rate of hot gas onto the SMBH follows a Bondi-Hoyle formulation \citep{bondi1952} and is only a function of the local temperature and the mass of the SMBH:

\begin{equation}
\rm{\dot{M}_{BH, R}=\kappa_{R}\dfrac{15}{16}\pi G~\overline{\mu}m_p\dfrac{kT}{\Lambda}M_{BH}}.
\label{equ:c16radio}
\end{equation}

In \Equ{c16radio}, $\rm{\kappa_{R}}$ is the ``radio-mode efficiency" parameter with a value of 0.08, $\rm \overline{\mu} m_p$ is the mean particle mass, $\rm T$ is the local temperature and $\rm \Lambda = \Lambda(T, Z)$ is the gas cooling function that depends on the temperature and the metallicity.
Using \Equ{lum_BH}, we can use the accretion rate to calculate the luminosity of the SMBH.
The accretion onto the black hole acts as a heating mechanism for the gas and the heating rate for the radio mode feedback can be quantified as

\begin{equation}
\rm{\dot{M}_{heat}=\dfrac{\dot{E}_{radio}}{0.5V_{vir}^2},}
\end{equation}

where the numerator is the luminosity of the black hole given by \Equ{lum_BH} and the denominator represents the specific energy of the gas in the halo.

The biggest update on the SAM from \cite{croton06} is the coarse coupling of the heating and cooling mechanism of the halo gas.
\citetalias{croton16} defines a heating radius, $\rm{R_{heat}}$, inside which the gas never cools.
At this radius, the energy injected by radio mode feedback is equal to the energy the halo gas would loose to cool onto the galaxy disk.
In this coupled heating-cooling cycle, the cooling rate of gas becomes

\begin{equation}
\rm{\dot{M}_{cool}'=\Bigg(1-\dfrac{R_{heat}}{R_{cool}}\Bigg)\dot{M}_{cool}.}
\label{equ:sage_cooling}
\end{equation}

In \Equ{sage_cooling}, $\rm{R_{cool}}$ is the cooling radius such that

\begin{equation}
\rm{\dot{M}_{cool}=\frac{1}{2}\Bigg(\dfrac{R_{cool}}{R_{vir}} \Bigg)\Bigg(\dfrac{M_{hot}}{t_{cool}} \Bigg)}.
\label{equ:sage_coolingrate}
\end{equation}

In this case, gas can only cool between $\rm{R_{heat}}$ and $\rm{R_{cool}}$ and if $\rm{R_{heat} > R_{cool}}$, cooling of gas is quenched.
The heating radius in the model is only allowed to increase in size in order to retain memory of previous heating episodes.

\subsection{Choice of SAMs}

The two SAMs we employ in this study are representative of the different codes used in the literature (see e.g \citealp{knebe18} for a comparison between different SAMs) and their predictions are easily available through web interface.
  Moreover, these two SAMs are well suited for our purposes, i.e. quantifying the impact of radio-mode AGN feedback on the onset of the passive fraction of observed galaxies.

These two codes, although both starting from the original \cite{croton06}, represent quite different approaches to the implementation of radio-model AGN feedback.
  On the one hand, \hen still employs a phenomenological prescription whose main aim is to reproduce the high-mass end of the SMF by quenching the cooling flows expected at the centre of massive dark matter halos.
  In \citetalias{henriques14}, there is no attempt to model the details of the gas accretion onto the central SMBH, but the main dependencies of radio-mode luminosities as a function of macroscopic quantities such as the hot gas and/or the SMBH mass.
  A number of different SAM codes share the same approach (like \citealp{Bower2006}, \citealp{DLB07} and \citealp{guo2011}).

On the other hand, \sage employs a more physical approach to gas accretion, trying to account for the detailed physics of the coupled gas cooling-heating cycle.
  The \citetalias{croton16} model is representative of an approach that has been considered (although in different frameworks and with different levels of sophistication) also by \cite{Monaco2007}, \cite{somerville08} and \cite{Fanidakis2011}.

Therefore, the comparison between these two models provides us with valuable insights on the effect of these two approaches on the overall galaxy population, while keeping the number of models to a manageable number.


\section{Observational Data}

\begin{figure*}
\begin{multicols}{2}
    \includegraphics[width=\linewidth]{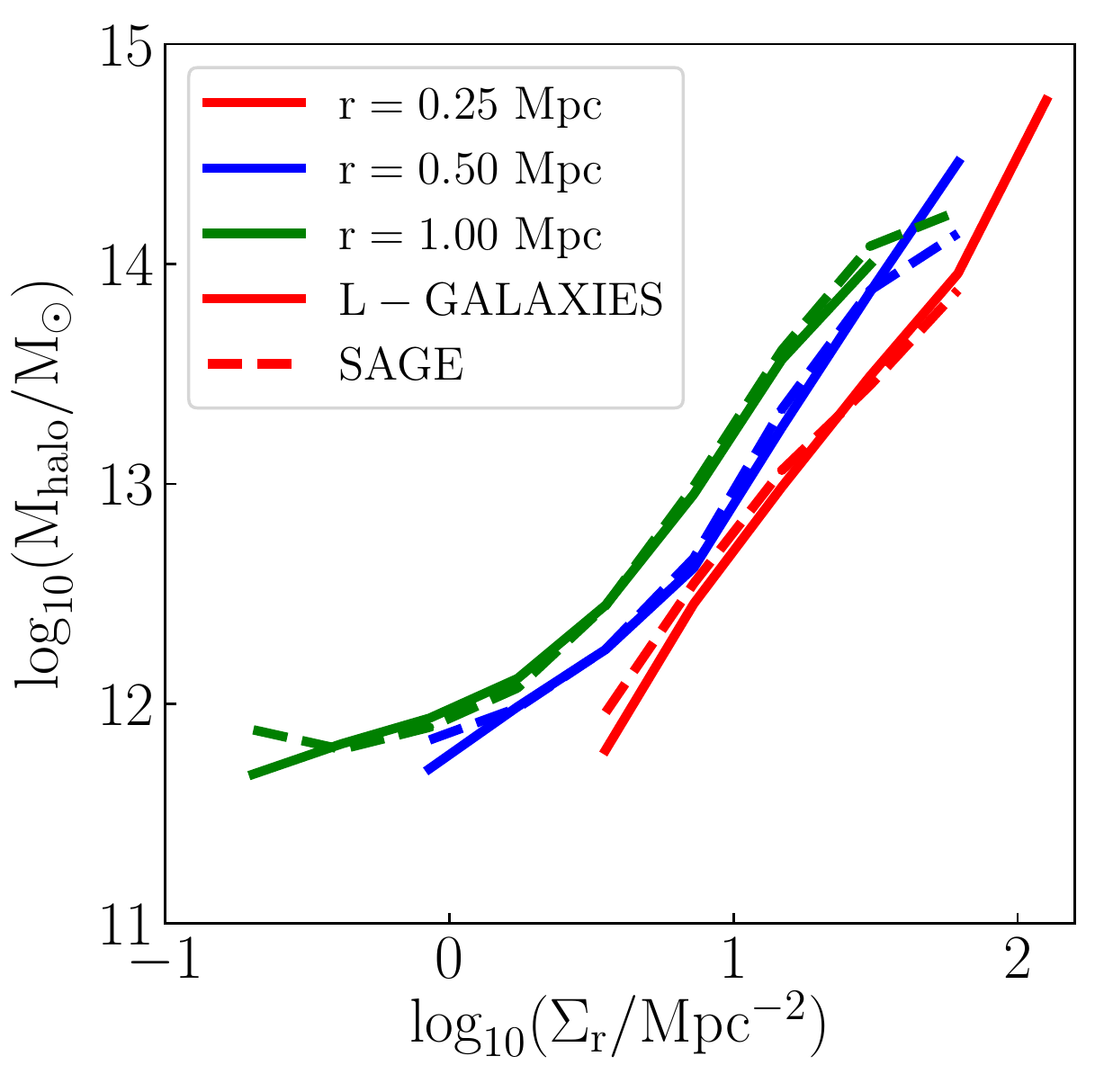}\par 
    \includegraphics[width=\linewidth]{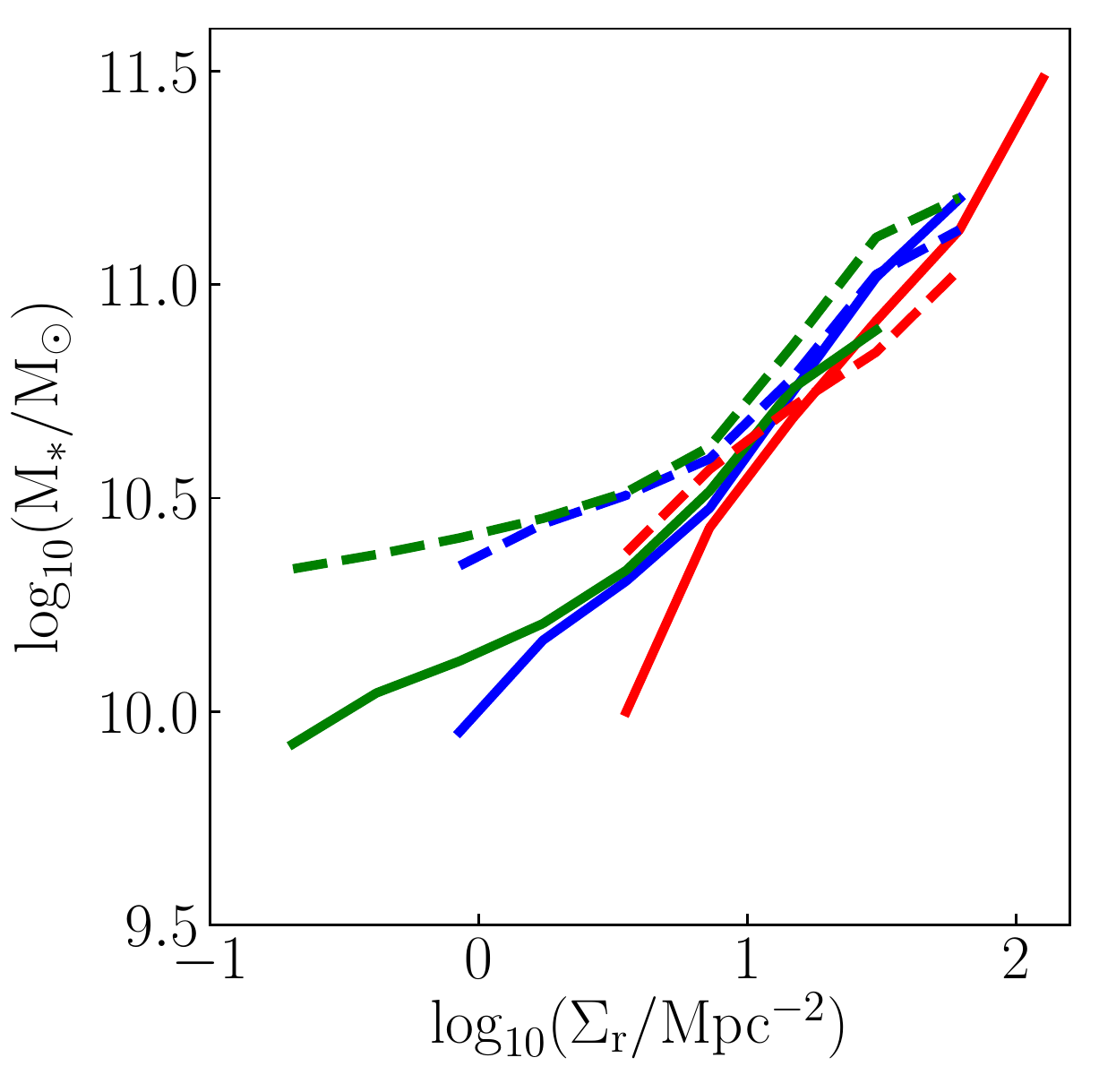}\par 
    \end{multicols}
\caption{Median dark matter halo mass (left panel) and stellar mass (right panel) as a function of neighbour density for various aperture sizes (different colours) at $z=0.0$.
  The solid lines show central galaxies from {\scriptsize L-GALAXIES} and dashed lines show central galaxies from \sage.}
\label{fig:mass_dens}
\end{figure*}

For comparison to our galaxy formation models, we use a SDSS-DR7 \citep{sdss2009} dataset.
The observed data is build with a modified catalogue from \cite{wilman10} which is drawn from the SDSS-DR7 sample.
The datasets provides the number of neighbours for every primary galaxy in a cylindrical aperture of different projected radii ranging between $0.1-3.0\rm{Mpc}$.
To make sure that the sample is volume complete, we limit the dataset to r-band absolute magnitude of $M_r\leq -20$ and a depth of $z\leq0.08$.
Furthermore, in order to account for the `missing galaxies' due to Malmquist bias, each galaxy is assigned a weight that is the ratio between the maximum volume in which these galaxies could be observed over the volume of the whole sample.
Using these weights, we calculate the passive fraction within a bin is calculated as

\begin{equation}
\rm{f_{pass}=\dfrac{\Sigma w_{pass}}{\Sigma w_{all}}}.
\end{equation}

In order to select passive galaxies, we use specific star formation rates defined as $\rm sSFR=SFR/M_{*}$.
A systems is defined to be passive if $\rm sSFR < 0.3t_{hubble}^{-1}\approx 10^{-11}yr^{-1}$ \citep{Franx2008, hirschmann14}.
The stellar masses and star formation rates are obtained by cross-correlating the sample with the MPA-JHU catalogue \citep{kauffmann03, 
Brinchmann04}.


\section{Selection of the Parameter Space and central galaxies}

For a fair comparison between the observed and simulated galaxies, with respect to the star formation suppression, an observationally motivated parameter space is critical.
In the ideal case, star formation quenching should be studied in the stellar mass - halo mass parameter space.
Properties such as star formation rates, optical sizes of galaxies are sensitive to its own growth history and therefore correlate strongly with stellar mass.
Similarly, dynamic properties such as maximum circular velocity, virial velocity scale with the dark matter halo mass.
However, using the stellar mass - halo mass parameter space comes with two disadvantages.
Firstly, the strong degeneracy between stellar mass and halo mass \citep{Yang2008} make it difficult to decide which parameter dominates passive fractions.
Furthermore, estimating the dark matter halo mass for a large number of galaxies can only be done indirectly and is accompanied by large uncertainties.

Both obstacles can be avoided by using a parameter that is observationally motivated and correlated with halo mass.
For this study, we analyze the passive fractions in an observationally motivated parameter space of stellar mass - neighbour density.
The neighbour density is a measure of the local environment and correlates strongly with halo mass.
We calculate the neighbour density for our model galaxies in fashion similar to \cite{wilman10} and \cite{fossati2015}. This allows us to study the impact of environment on various galaxy properties.
For a galaxy, the neighbour density is calculated to be

\begin{equation}
\Sigma_R = \dfrac{N_{R}}{\pi R^2}.
\label{equ:density}
\end{equation}

These densities are calculated within a projected aperture at various radii, $R$, ranging from $0.1-3.0\rm{Mpc}$.
Galaxies are counted as neighbour if they fall within the said aperture and their Hubble flow velocities are within a velocity width of $dv=\pm \rm{1500~km~s^{-1}}$.
In this framework, galaxies with $N_{R}=0$ are considered isolated systems.
Isolated systems are given a arbitrary value that is equal to the half the minimum density for galaxies that have neighbours.
To ensure they are represented on a logarithmic scale.

\Fig{mass_dens} shows the median halo mass (left panel) as a function of the density in apertures of various projected radii (various colours) for central galaxies as defined from the simulation.
The two different galaxy formation models are shown using different line style.
The strong correlation between environment and halo mass is seen for both SAMs which is due to the fact that both were run on the millennium simulation.
Lower density environment trace lower halo masses. This expected as higher mass create a deeper gravitational potential which leads to a denser environment.
In the local Universe, halo mass as a function of environment is independent of the aperture size.

The right panel is figure \Fig{mass_dens} shows the median stellar mass of central galaxies as a function neighbour density for various aperture radii.
Like with halo mass, we notice a strong correlation between stellar mass and density.
Galaxies with large stellar masses live in denser environment.
At low densities, \sage produces higher stellar mass galaxies than {\scriptsize L-GALAXIES} by $\sim \rm 0.3~dex$ even though the median halo mass is comparable.
The reason to this higher stellar mass is suspected to be the merger rate and the treatment of the orphan galaxies discussed in \Sec{sage}.
Another reason for the disagreement might be a weak stellar feedback prescription allowing for the gas to cool down in low density environment to form stars.
However, a complete analysis of this problem is out of the scope of this study.

For selecting central galaxies from both models and observations, we use a modified scheme of mass ranks within a cylindrical aperture presented in \cite{fossati2015}.
We refer the interested readers to the \App{central_select} for more details in the adopted scheme and its differences with respect to \cite{fossati2015}.
The depth of the cylinder is selected in velocity space and the radius in calculated using \Equ{adaptive}.
For both SAMs and observations, we use the adaptive aperture with $\rm{n=8}$, $\rm{r_{max}=2.5~Mpc}$ and $\rm{v_{depth}=2000km~s^{-1}}$ to select central galaxies for the rest of the study.
The choice of these parameters provide a good balance between the completeness of the selections and its purity.


\begin{figure*}
\begin{multicols}{2}
    \includegraphics[width=\linewidth]{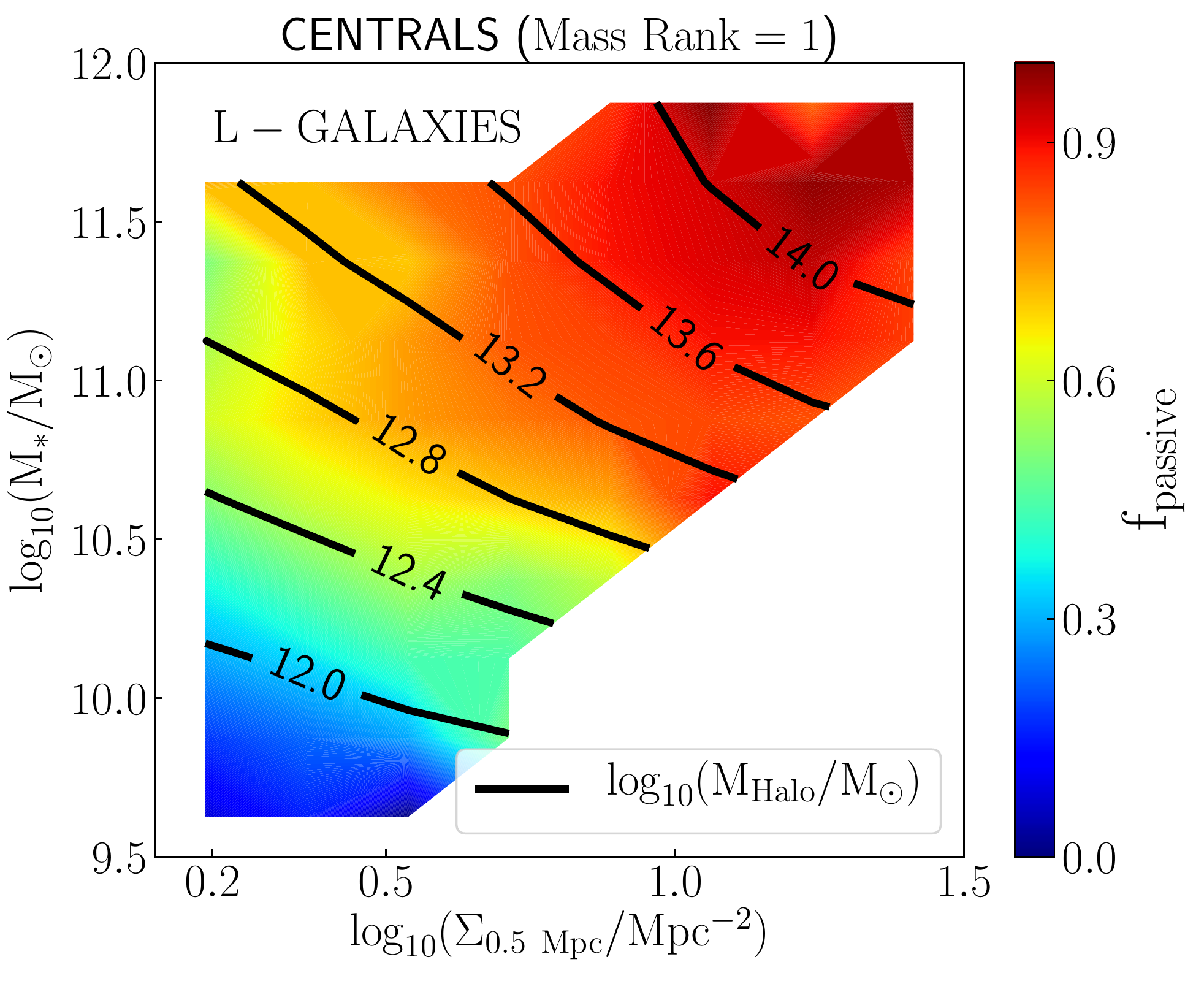}\par 
    \includegraphics[width=\linewidth]{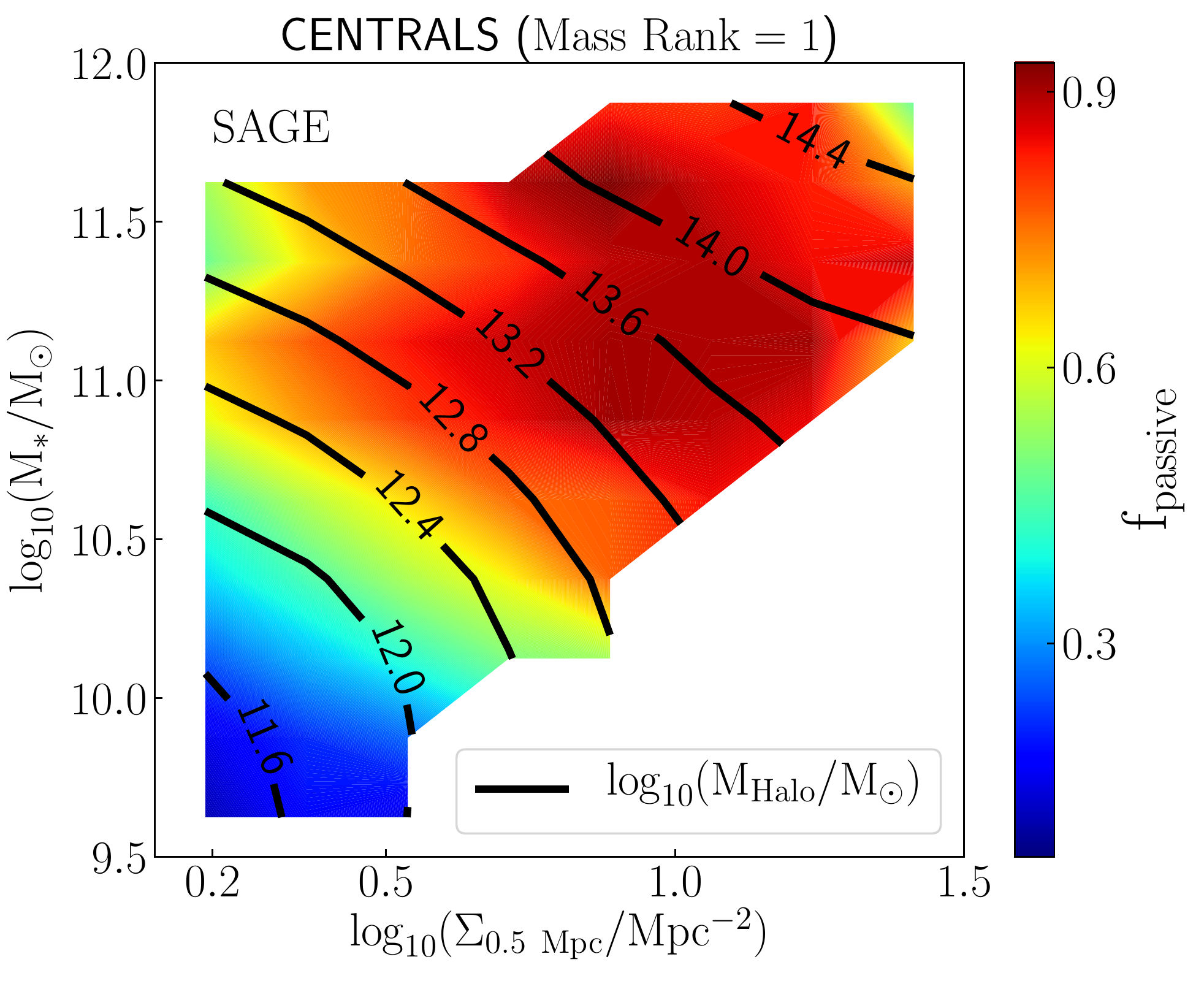}\par 
    \end{multicols}
\caption{Passive fractions in the $M_*-\Sigma_{r}$ parameter space for \hen (left panel) and \sage (right panel).  We use $0.5~\rm Mpc$ aperture for calculating the neighbour density for model galaxies. The passive fraction are shown for central galaxies that are selected by assigning mass ranks using the adaptive aperture $r(8, 2.5, 2000)$. The contours presented on both panels represents the median halo mass in each bin in log space.}
\label{fig:p_frac}
\end{figure*}

\begin{figure*}
\begin{multicols}{2}
    \includegraphics[width=\linewidth]{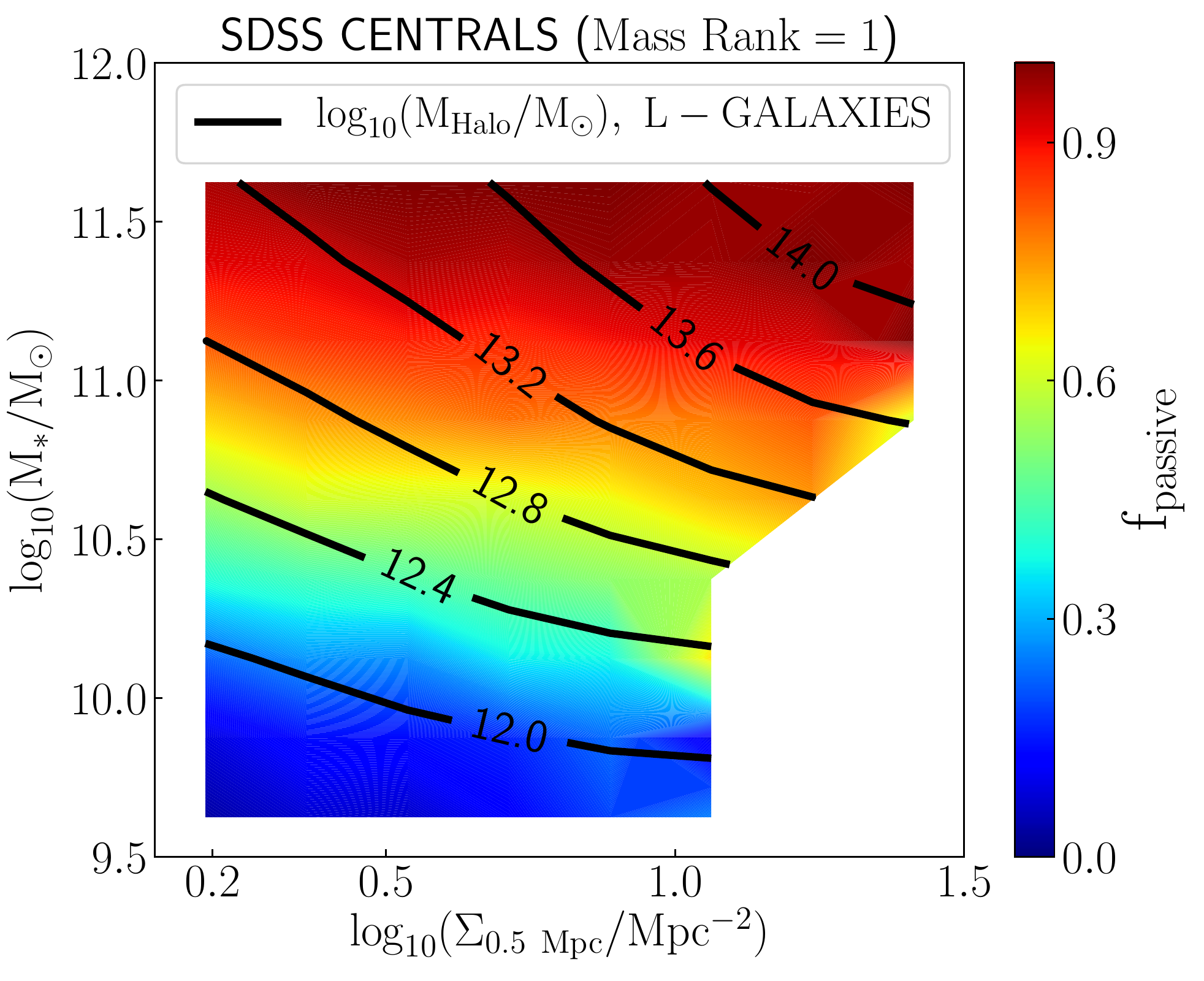}\par 
    \includegraphics[width=\linewidth]{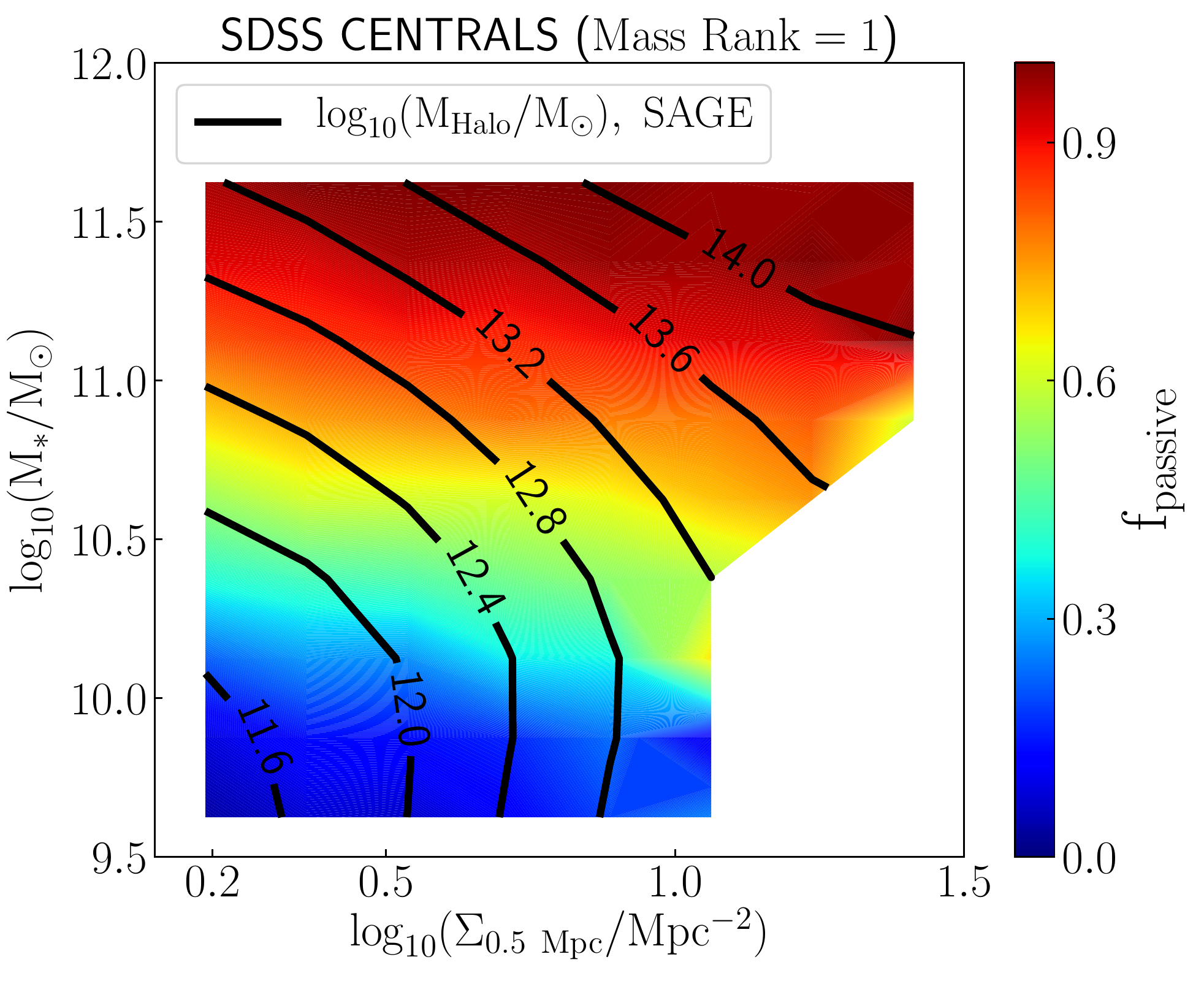}\par 
    \end{multicols}
\caption{Passive fractions in the $M_*-\Sigma_{r}$ parameter space SDSS central galaxies selected using mass ranks assigned by the adaptive aperture. We use $0.5~\rm Mpc$ aperture for calculating the neighbour density around central galaxies. The contours show the median halo mass for the two SAMs, \hen (left panel) and \sage (right panel).}
\label{fig:obs_p_frac}
\end{figure*}

\begin{figure*}
\begin{multicols}{2}
    \includegraphics[width=\linewidth]{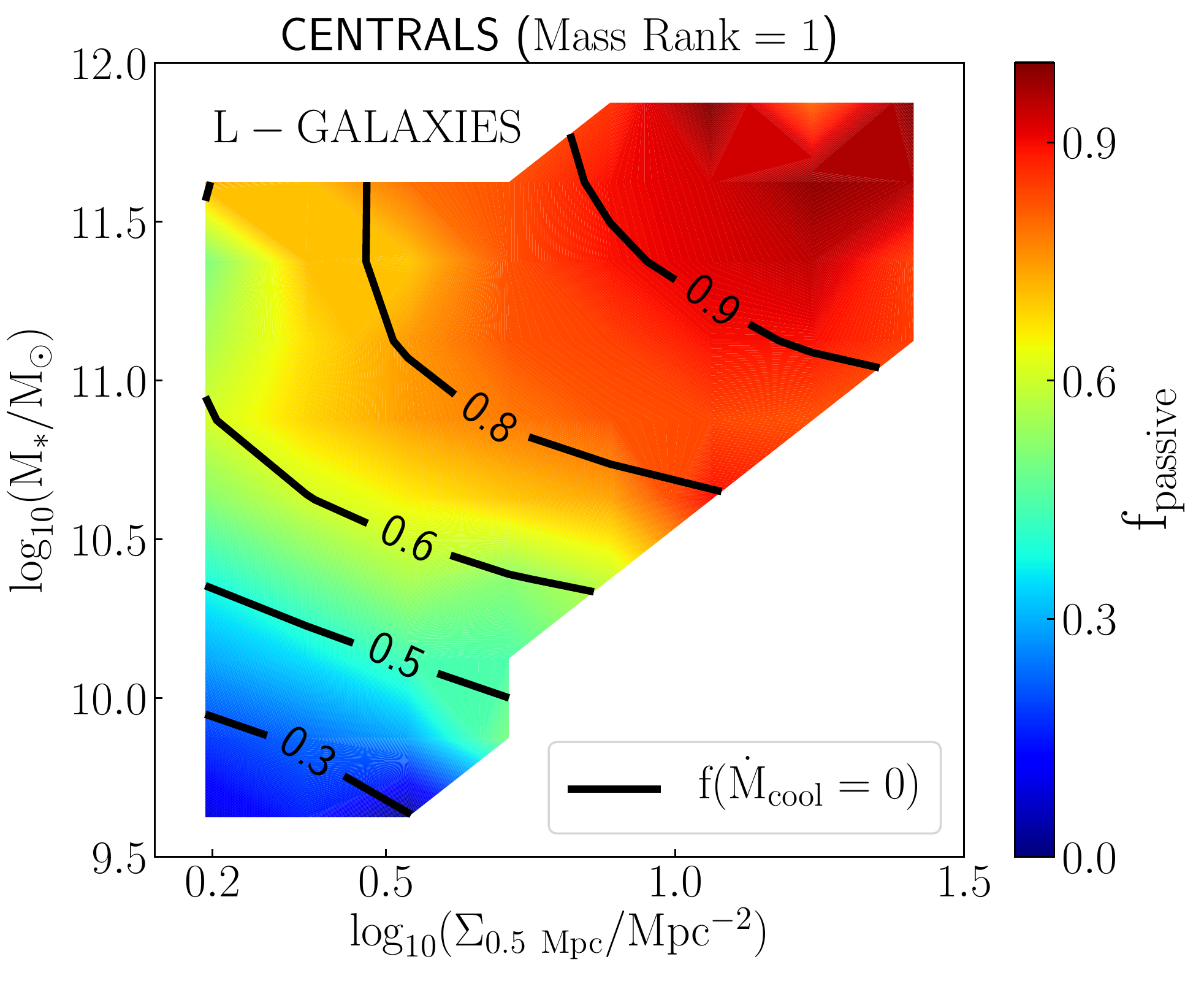}\par 
    \includegraphics[width=\linewidth]{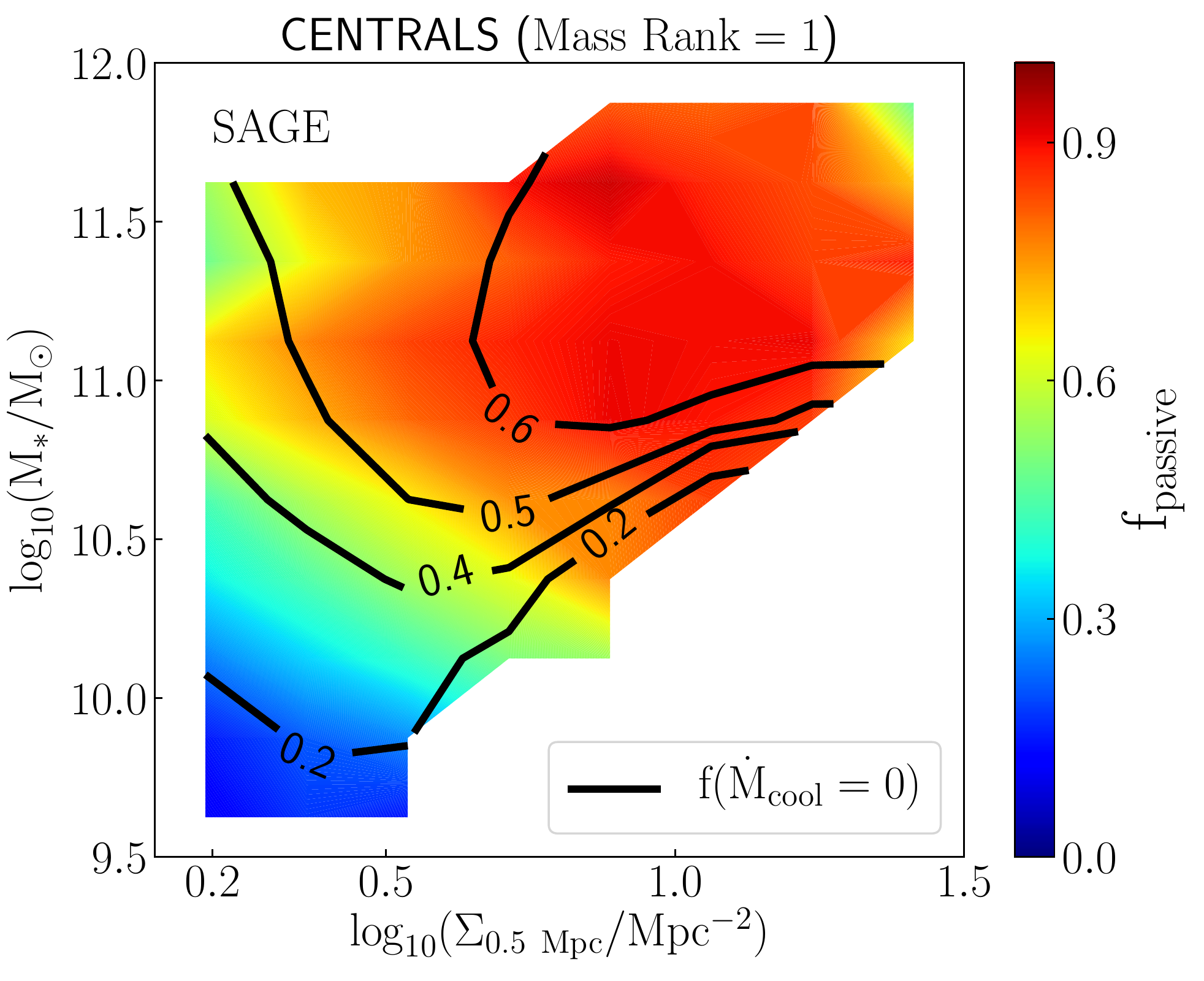}\par 
    \end{multicols}
\caption{Passive fractions in the $M_*-\Sigma_{r}$ parameter space for \hen (left panel) and \sage (right panel). We use $0.5~\rm Mpc$ aperture for calculating the neighbour density for model galaxies. The passive fraction are shown for central galaxies that are selected by assigning mass ranks using the adaptive aperture $r(8, 2.5, 2000)$. The contours on top show the fraction of galaxies where the cooling of hot gas due to radio mode AGN feedback is recorded to be zero.}
\label{fig:den_hot_cooling}
\end{figure*}


\section{Passive Fractions in Galaxies}


\subsection{Dependence on Halo Mass}

We start with a discussion about the behaviour of passive fraction for central galaxies as a function of stellar mass, halo mass and galaxy density (\Fig{p_frac}).
We see that most star forming galaxies reside in low density environments and have lower stellar/halo masses.
Similarly, most quenched galaxies live in high density environment like groups or clusters are typically very massive in stellar or halo mass.
Furthermore, a relation between passive fraction and halo mass is also seen.
Galaxies living in small halos are actively star forming whereas those living in massive halos seem to be passive. Qualitatively, these trends are the same for both \hen and \sage galaxy formation models.
An interesting feature of \Fig{p_frac} is the lack galaxies in high density and low mass $\rm{(\log(M_*/M_{\odot})<10,~\log(\Sigma_{0.5Mpc}/Mpc^{-2})>0.5)}$.
  Due to the low number of galaxies in this part of the parameter space, we choose to remove these systems, in order to avoid spurious conclusions due to low-number statistics.
  In order to prevent that, we considered bins of stellar mass and galaxy density with more than 30 objects in them.

We next superimpose the halo mass contours shown in \Fig{p_frac} to the passive fraction from the observed SDSS data in \Fig{obs_p_frac}.
The left panel represents the contours from \hen while \sage prediction are shown on the right.
In the left panel of \Fig{obs_p_frac}, we notice that the observed  star forming central galaxies living in isolated environment are hosted in low halo masses.
On the high mass, high density region in the parameter space, trends similar to model passive fraction are observed.
The most massive galaxies in the local Universe are quenched regardless of the environment that they live in.
Star formation quenching for observed central galaxies appears to be mainly driven by a quantity that correlates with stellar masses.
Although the contours do not match the SDSS data perfectly, \hen seems to catch the dominant dependence of quenching on stellar mass while \sage predicts a density/halo mass to play a significant role, bigger than what SDSS is telling us.

An interesting feature seen in \Fig{obs_p_frac} are the very massive quenched galaxies that live in low density environment which are not reproduced by \hen and \sage galaxy formation models.


\subsection{Dependence on AGN Heating}

\begin{figure*}
\begin{multicols}{2}
    \includegraphics[width=\linewidth]{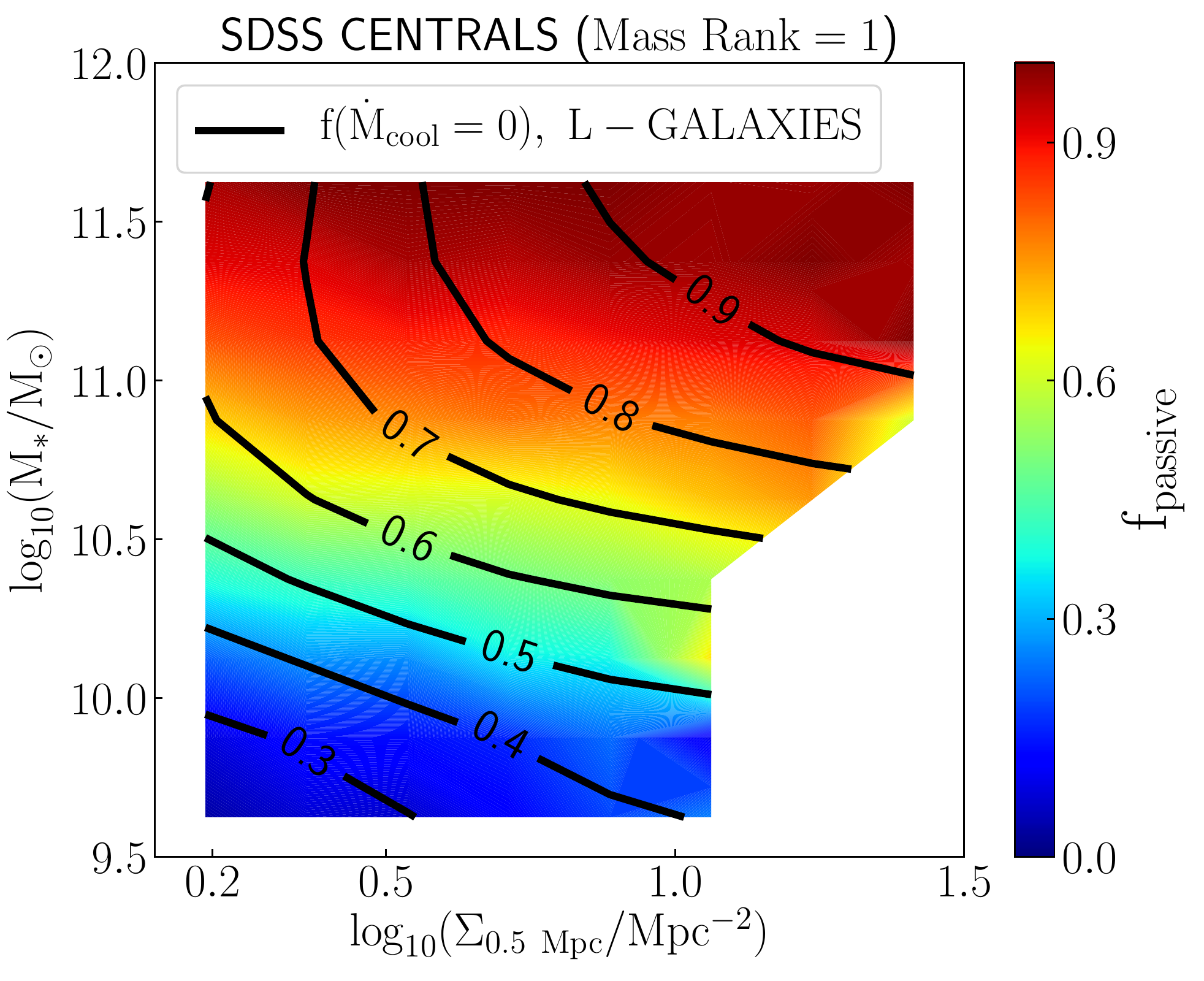}\par 
    \includegraphics[width=\linewidth]{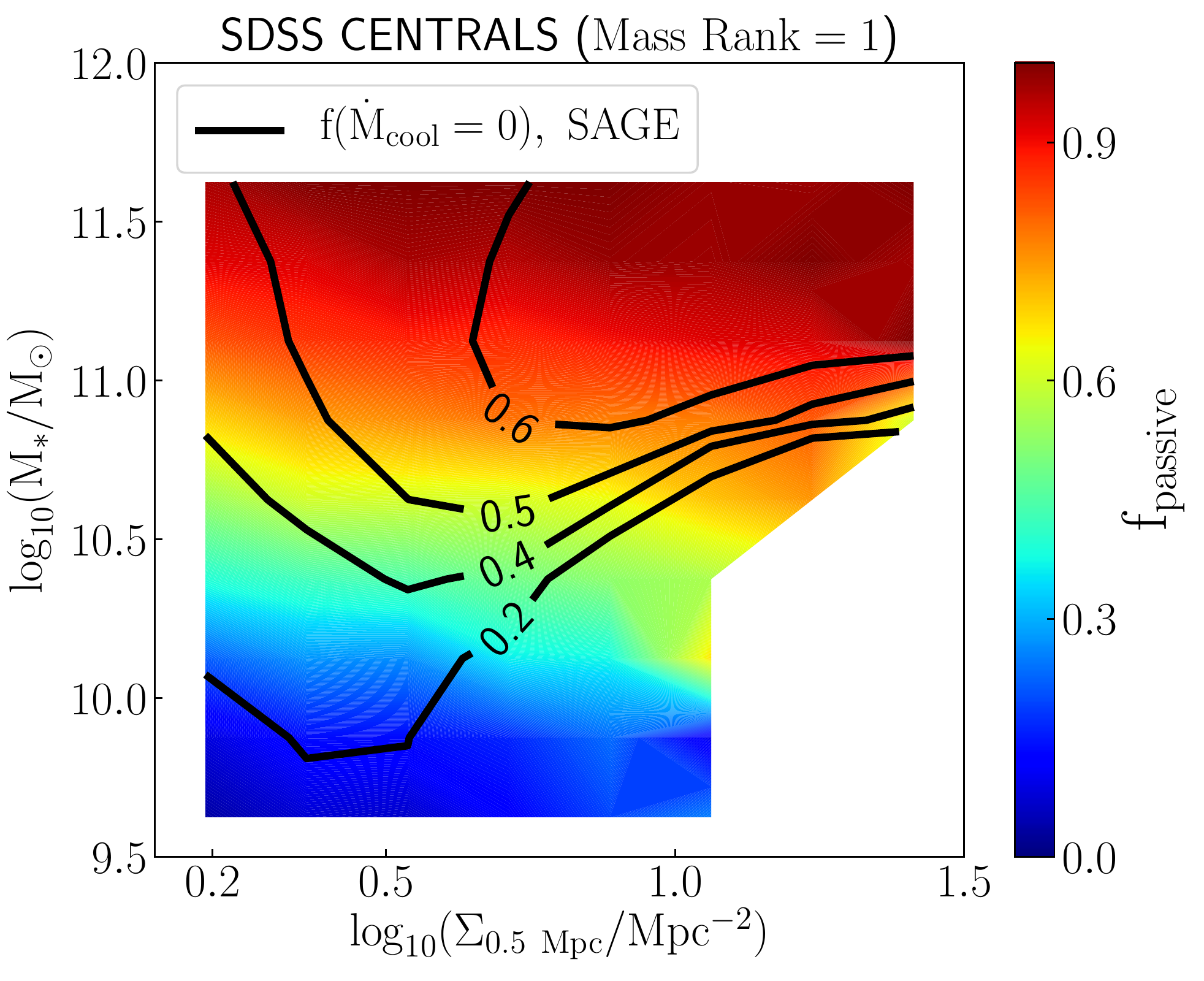}\par 
    \end{multicols}
\caption{Passive fractions in the $M_*-\Sigma_{r}$ parameter space SDSS central galaxies selected using mass ranks assigned by the adaptive aperture. We use $0.5~\rm Mpc$ aperture for calculating the neighbour density around central galaxies. The contours on top show the fraction of galaxies where the cooling of hot gas due to radio mode AGN feedback is recorded to be zero, \hen (left panel) and \sage (right panel).}
\label{fig:obs_hot_cooling}
\end{figure*}

\begin{figure}
\includegraphics[scale=0.45]{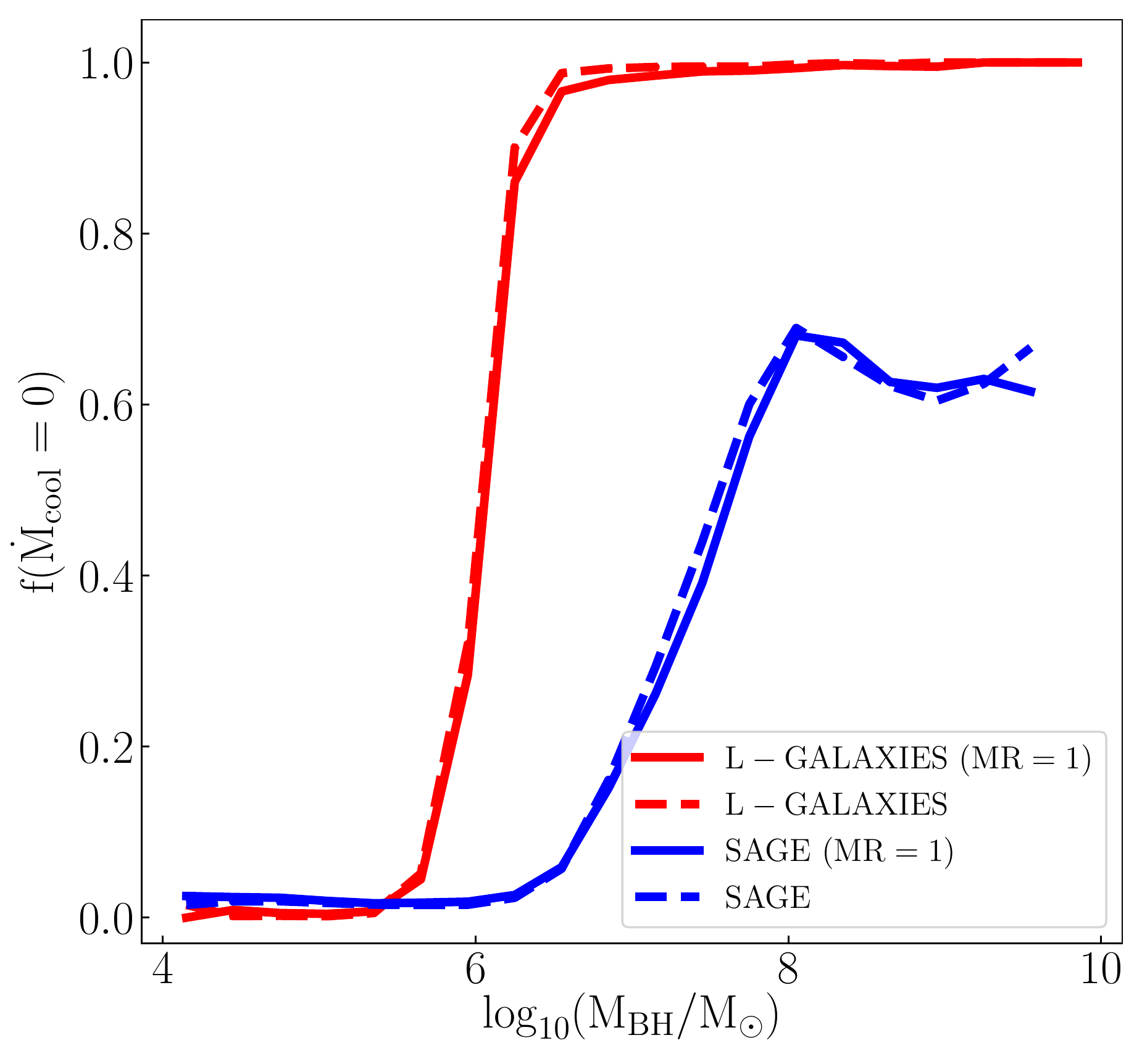}
\caption{Fraction of galaxies with zero gas cooling rate as a function of black hole mass. The central galaxies selected using mass ranks assigned by the adaptive aperture are presented using solid line and dashed line show the central galaxies as defined by the SAMs. Red represents \hen and blue represents \sage galaxies. The central galaxies defined by the SAMs and using our algorithm perform almost identically.}
\label{fig:halo_cooling}
\end{figure}

\begin{figure*}
\begin{multicols}{2}
    \includegraphics[width=\linewidth]{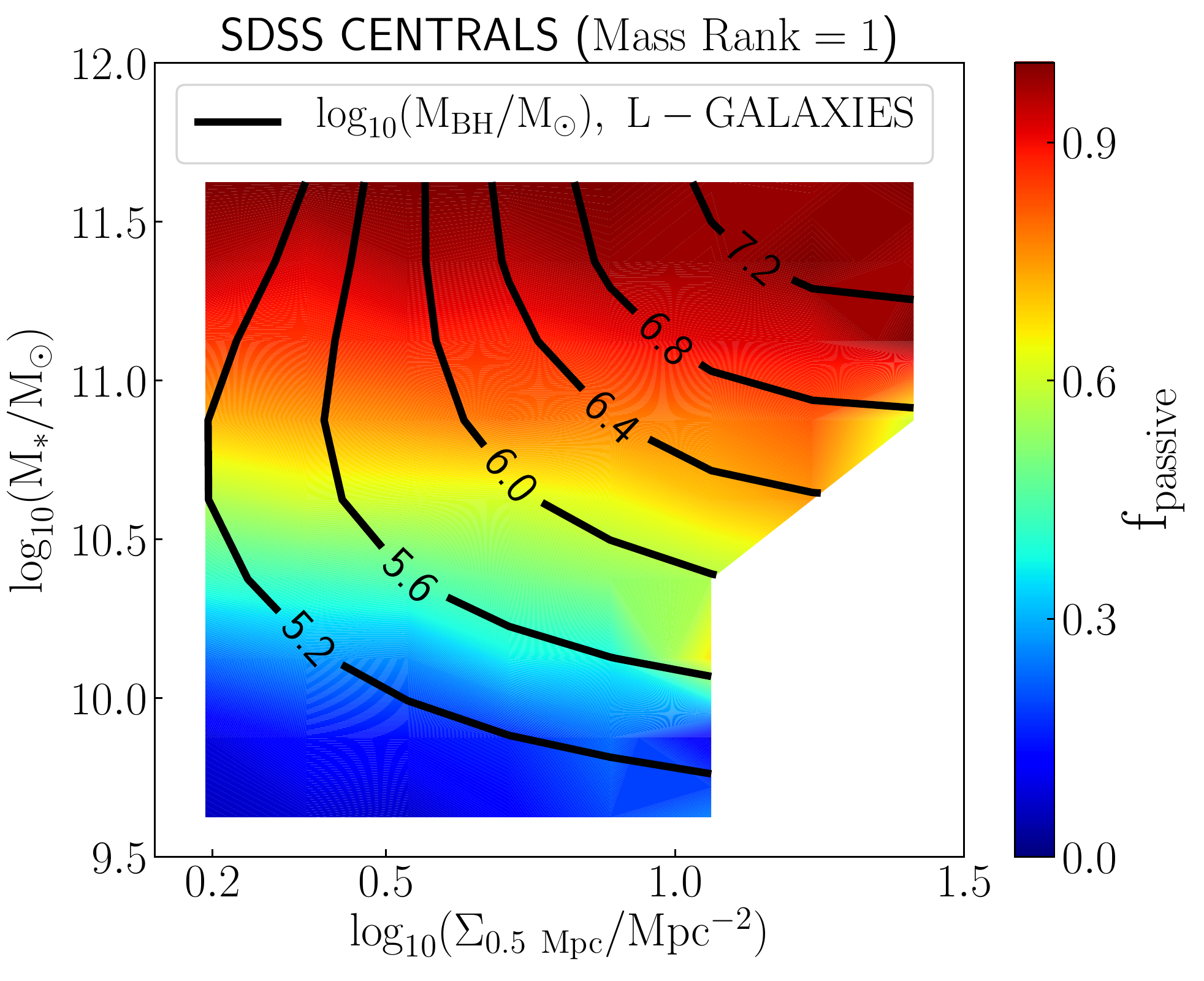}\par 
    \includegraphics[width=\linewidth]{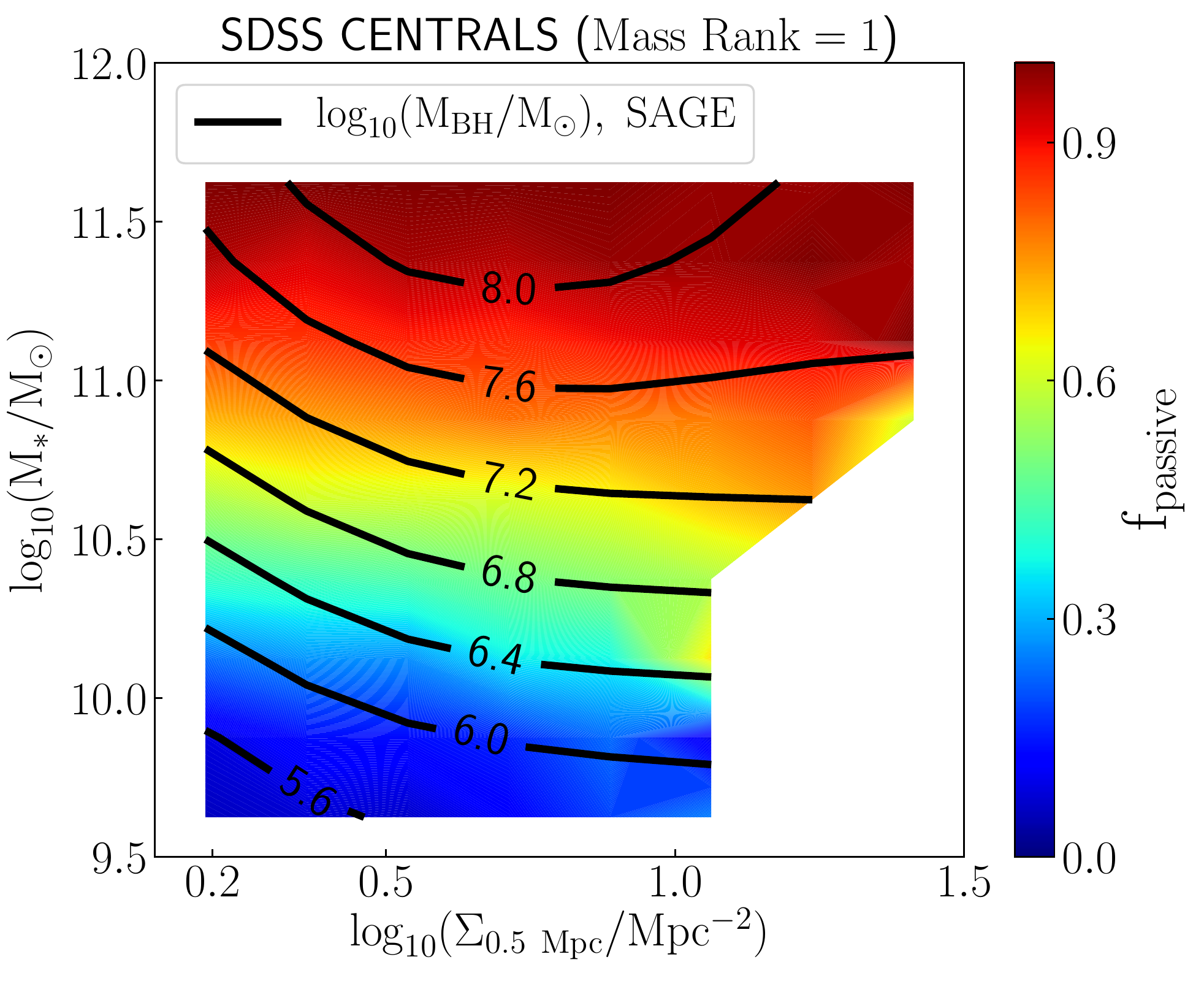}\par 
    \end{multicols}
\caption{Passive fractions in the $M_*-\Sigma_{r}$ parameter space SDSS central galaxies selected using mass ranks assigned by the adaptive aperture. We use $0.5~\rm Mpc$ aperture for calculating the neighbour density around central galaxies. The contours on top show the median black hole masses for \hen (left panel) and \sage (right panel).}
\label{fig:obs_bh}
\end{figure*}

For central galaxies, radio-mode AGN feedback would inject energy into the galaxies heating the gas, ultimately preventing cooling and subsequent gravitational collapse and ceasing star formation.
For both SAMs, we can calculate the rate of cooling of the hot gas in the presence of AGN radio mode feedback (see \Equ{cooling_rate} and \ref{equ:sage_cooling}).
\Fig{den_hot_cooling} presents the passive fraction for the two models in the $M_*-\Sigma_{r}$ parameter space with the contours representing the fraction of galaxies where gas cooling is prevented by feedback.
With \hen , the passive fraction corresponds closely to the fraction of galaxies with no cooling.
The right panel of \Fig{den_hot_cooling} shows the same plot for \sage.
While the quenched region of the parameter space still corresponds to maximum $\rm{f(\dot{M}_{cool}=0)}$, the fraction of galaxies with no cooling peaks at $\sim 60\%$.
The smaller value for $\rm{f(\dot{M}_{cool}=0)}$ is suggestive of a competition between the radio mode AGN feedback and formation of stars.
Furthermore, even though both SAMs use the same equation for star formation, the efficiency of converting gas to stars has been tuned very differently.

In general, we can say that AGN feedback is dominant in quenching star formation in central galaxies.
The region of high passive fraction in \Fig{den_hot_cooling} corresponds to galaxies which have little no cooling of gas.

\Fig{obs_hot_cooling} represents the passive fractions for the observed SDSS sample, for central galaxies as a function of stellar mass and density on an aperture of $0.5~\rm Mpc$.
The contours are the same as in \Fig{den_hot_cooling} for both \hen (left panel) and \sage (right panel).
We test the possibility that galaxies that are a part of massive halos should be passive due to presence of SMBHs that injects enough energy into the galaxy to suppress cooling leading to quenching of star formation. 

For \hen the impact of AGN feedback is not very significant for less massive and fairly isolated systems, only $30\%$ of the galaxies have gas cooling completely suppressed due to radio mode feedback.
For low mass galaxies, the fraction of galaxies with no cooling agrees very well with the fraction of passive galaxies.
With increasing stellar masses and galaxy density, the fraction of galaxies without gas cooling goes up to $90\%$. However, observed massive central galaxies with low star formation rate have varying $\rm{f(\dot{M}_{cool}=0)\simeq 0.7-0.9}$.
On the other hand, \sage contours in the right panel of \Fig{obs_hot_cooling} look more different that \hen and diverge from the observed passive fraction.
Even with the more complex radio mode AGN feedback, low $\rm{f(\dot{M}_{cool}=0)}$ in the passive region of the observed parameter space presents a challenge for \sage.


\subsection{Dependence on Black Hole Mass}

\begin{figure*}
\begin{multicols}{2}
    \includegraphics[width=\linewidth]{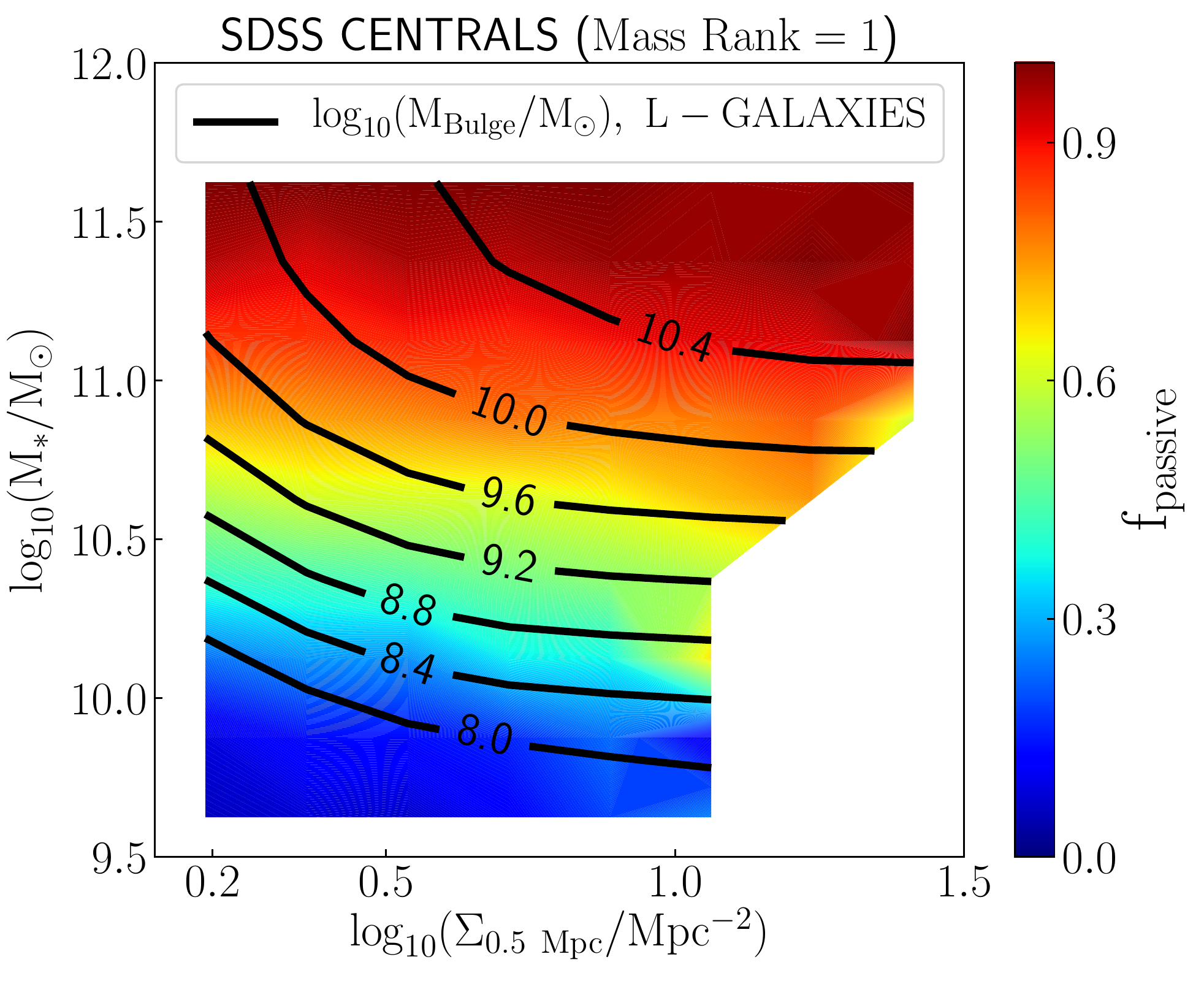}\par 
    \includegraphics[width=\linewidth]{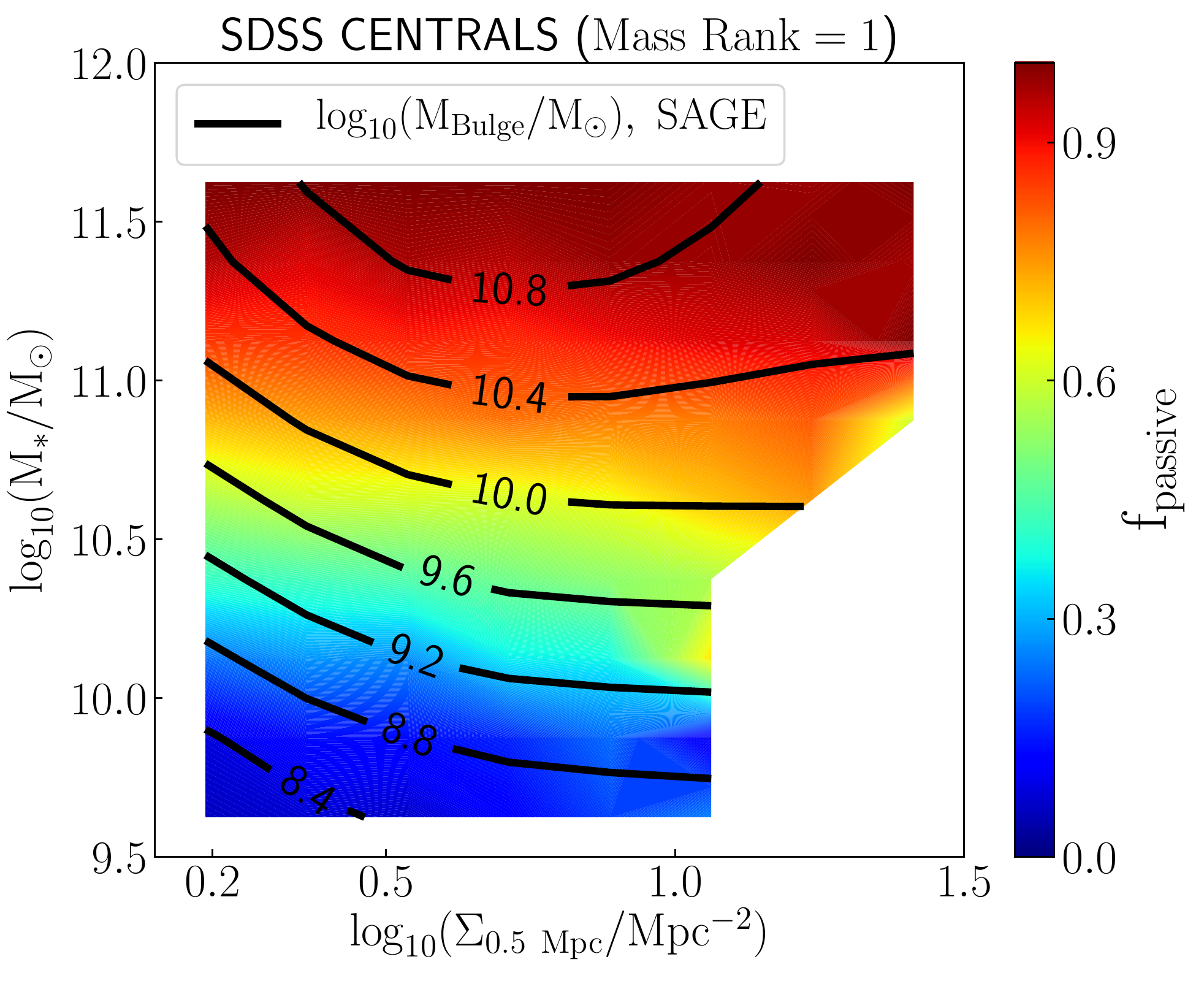}\par 
    \end{multicols}
\caption{Passive fractions in the $M_*-\Sigma_{r}$ parameter space SDSS central galaxies selected using mass ranks assigned by the adaptive aperture. We use $0.5~\rm Mpc$ aperture for calculating the neighbour density around central galaxies. The contours on top show the median bulge mass for \hen (left panel) and \sage (right panel).}
\label{fig:obs_bulge}
\end{figure*}

If the central black hole is responsible for quenching central galaxies at $\rm {z\sim0}$, then a relation black hole mass and the lack of cooling of gas should be seen in the model.
\Fig{halo_cooling} shows $\rm{f(\dot{M}_{cool}=0)}$ as function of $\rm{M_{BH}}$ for central galaxies selected using the mass rank assigned by the adaptive aperture for both SAMs.
AGN feedback in \hen very strongly controls the cooling of gas.
Nearly all central galaxies that host an SMBH with $\rm{\log_{10}(M_{BH}/M_{\odot})\geq 6.0}$ are extremely inefficient at cooling gas.
For \sage, gas cooling only starts to become inefficient for central galaxies with more massive SMBH ($\rm{\log_{10}(M_{BH}/M_{\odot})\sim 8.0}$) which corresponds to $\sim 60\%$ with no gas cooling.
A central galaxies population with $\rm{\log_{10}(M_{BH}/M_{\odot})>8.0}$ contains a significant number of galaxies that are cooling gas that could be still forming stars.
Cooling flows are expected to be the main fuel for star formation in these massive galaxies living in dense environments; however, gas rich mergers with gas-rich satellites can also provide cold gas and cause star formation.

\Fig{obs_bh} shows the observed passive fractions from \Fig{obs_hot_cooling} with overlying contours of median black hole mass for central galaxies in both SAMs.
For both simulations, the most massive black holes correspond to the passive part of the observed parameter space.
However, observed massive galaxies that live in isolated environments are passive correspond to low mass SMBH from \hen.
The high passive fractions observed for the massive isolated galaxies represents a possible tension with the AGN feedback scheme adopted in \hen, due to the predicted low mass of corresponding central SMBH.
Nonetheless, other causes, like the lack of cold gas, may explain star formation quenching in these objects.
This tension suggests a possible problem for the growth SMBH residing in central galaxies, especially in isolated environments in SAMs.

Qualitatively, the contours for SMBH mass for \sage show a better agreement with the observed passive fraction (see right panel \Fig{obs_bh}).
Independent of environment, massive centrals hosting massive SMBH correspond massive observed passive galaxies.
Despite the more massive SMBH, the fraction of centrals with suppressed gas cooling (see \Fig{obs_hot_cooling} and \ref{fig:halo_cooling}) seems to stay low, indicating inefficient heating.


\subsection{Dependence on Bulge Mass}

\begin{figure}

  \includegraphics[width=\linewidth]{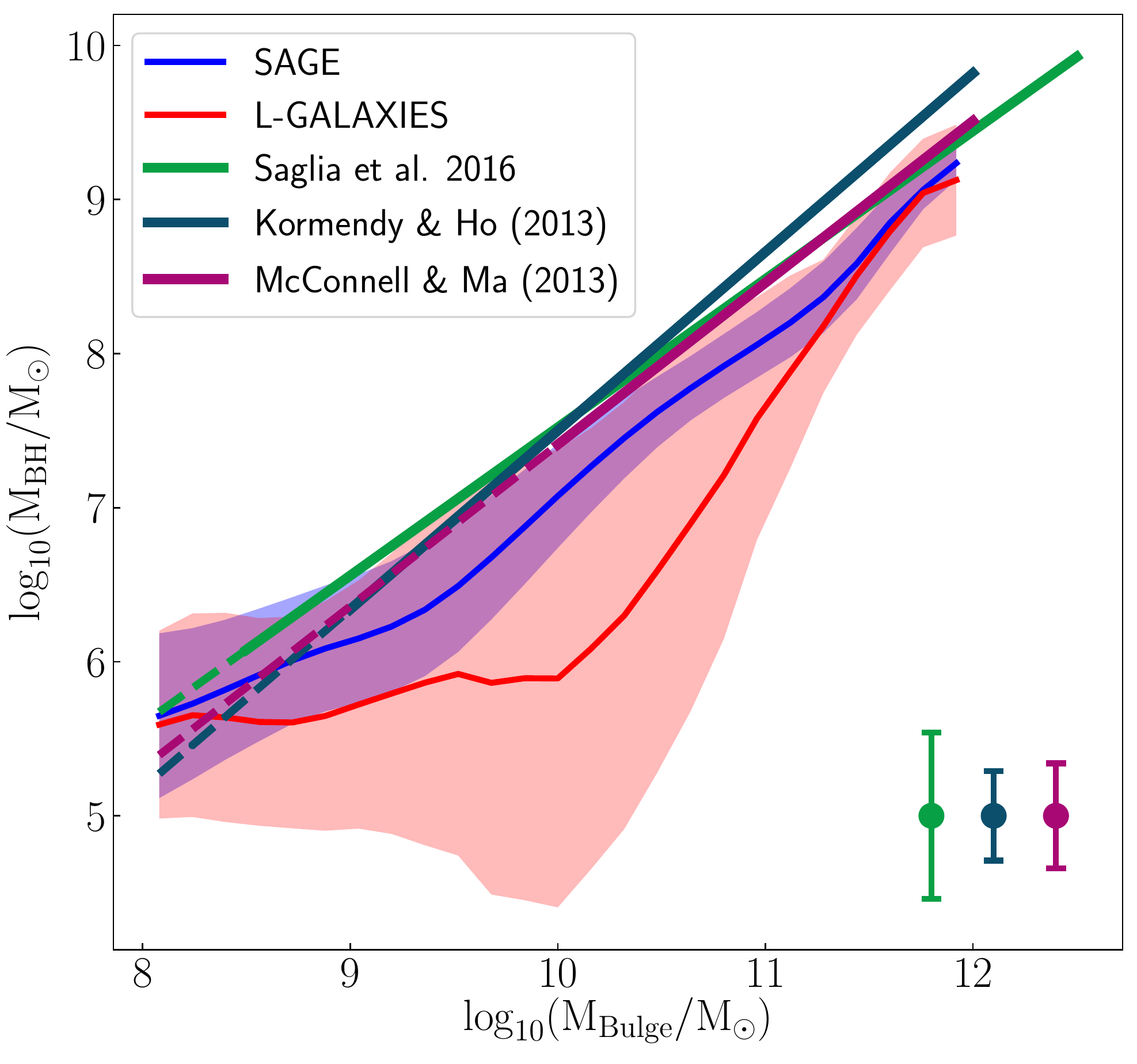}
  \caption{Black hole mass vs. bulge mass for the SAMs compared to various observed relation. For both SAMs, the line represent the mean calculated in a bulge mass bin of $\rm 0.16\,dex$ and the shaded region shows the standard deviation within the same bin. The left panel shows the central galaxies selected by the SAMs and the right panel shows the central galaxies selected using the adaptive aperture. The observed relation are best fit lines from \protect\cite{saglia16}, \protect\cite{Kormendy13} and \protect\cite{McConnell13}. The dashed lines for the observed relations show the extrapolation to a lower bulge mass range. The points shown on the lower right corner represent the scatter for the observed relations.}
  \label{fig:model_bubh}
\end{figure}

To explore star formation quenching for central galaxies, we next explore the role of bulge mass in SAMs.
The left panel in \Fig{obs_bulge} shows the observed passive fractions with overlying contours presenting the bulge mass for central galaxies in \hen.
In general, the observed star forming galaxies correspond to low bulge mass whereas passive observed galaxies host massive bulges. 

The observed passive fraction agrees strongly with the model bulge masses for both SAMs.
Active galaxies, independent of the environment, host smaller bulges.
As the fraction of quenched galaxies increases in the parameter space, so does the model bulge mass.
The disagreement between model and observed massive, quenched isolated galaxies could be studies using \Fig{obs_bulge}.
The left panel show the model bulge mass from \hen.
It is seen that observed massive, quenched isolated galaxies correspond to massive bulge and low mass black hole.
The reason for disagreement could be due to a large scatter in the $\rm{M_{Bulge}-M_{BH}}$ relation.

Looking at \Fig{obs_bh} and \ref{fig:obs_bulge}, we notice that both \sage has more massive black hole and bulges that \hen.
Such a situation is a direct consequence of the higher merger rate of galaxies in \sage than in \hen.
This implies that bulges grow much more efficiently and more mass is locked into bulges than is required \citep{knebe18}.
This merger excess might also help in making the SMBH more massive and might result in a tighter $\rm{M_{Bulge}-M_{BH}}$ relation \citep{Jahnke11}.
However, \sage still shows some net cooling in these galaxies which could indicate that gas cooling in \sage might be inconsistent with observations.
Comparing the two SAMs, we can conclude that in \hen quenching is mainly related to the halo and bulge mass and is more efficient.
In \sage, quenching is mainly related to the SMBH and bulge mass.
Neither SAM is able to completely capture the whole complexity of the observed passive galaxy population.


\begin{figure*}
\begin{multicols}{2}
    \includegraphics[width=\linewidth]{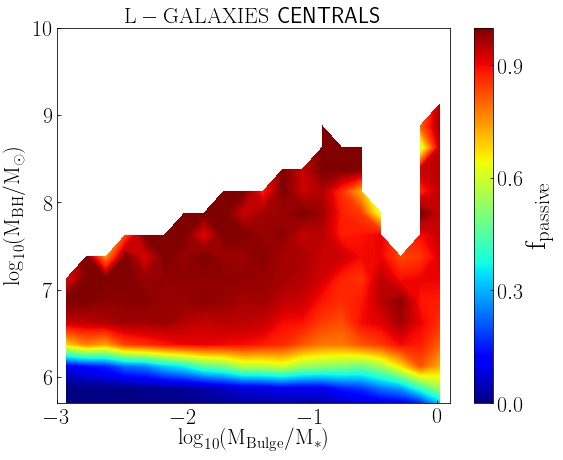}\par 
    \includegraphics[width=\linewidth]{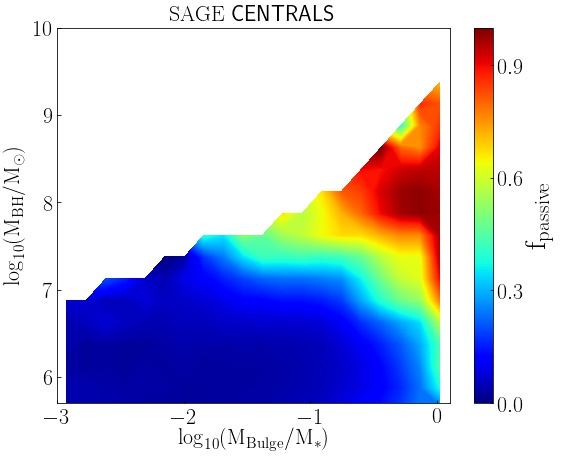}\par 
    \end{multicols}
\caption{Passive fractions in the $\rm  M_{BH}-M_{Bulge}/M_{*}$ parameter space for central galaxies in \hen (left panel) and \sage (right panel).}
\label{fig:passive_bubh}
\end{figure*}

\section{Black Hole - Bulge Mass relation}

In the previous sections, we have shown that star formation quenching seems to be driven either by the presence of an AGN or a massive bulge.
Central galaxies from \hen and \sage show that the presence of massive SMBH can suppress the gas cooling and eventually cease star formation.
In general, the star formation activity in central galaxies is driven by the coupling between the black hole and bulge mass and their growth mechanisms.
Various studies, observational and theoretical, have reported a tight relationship between the mass of the central black hole and the mass of the bulge \citep{merritt2000, haring04, croton2006b, allesandra2012, saglia16}.
In this section, we discuss the black hole - bulge mass relation for the SAMs used in this study.
\Fig{model_bubh} shows the $\rm{M_{BH}-M_{Bulge}}$ relation for \hen and \sage centrals and comparisons with best fit power laws provided in \cite{saglia16}, \cite{Kormendy13} and \cite{McConnell13} (solid straight lines).
The red and blue solid lines show the mean black hole mass for \hen and \sage respectively.
The shaded region represents the standard deviation within the same bulge mass bin for the SAMs, while the points with error bars in the lower right corner show the typical scatter in the observed studies.

\Fig{model_bubh} shows the two models predict quite a different shape of the $\rm{M_{BH}-M_{Bulge}}$ relation.
  In general, both models predict SMBH masses at fixed $\rm{M_{Bulge}}$ that are lower than observational estimates, but that agree on a $\rm{1-\sigma}$ level.
  \sage predicts an almost constant scatter with bulge mass, and much smaller than \hen.
  On average the average standard deviation in \hen is $\rm 0.56\, dex$ higher than \sage.

In detail, $\rm {M_{BH}}$ are always under-predicted in the range $\rm 8.5 \lesssim \log_{10}(M_{Bulge}/M_{\odot}) \lesssim 11.0$, with better agreement at higher and lower bulge masses.
  \sage predicts a slope for the $\rm{M_{BH}-M_{Bulge}}$ relation that is closer to the observation estimate (due to the fact that the model has been explicitly calibrated to reproduce the constrain).
  On the other hand, \hen strongly under-predicts $\rm {M_{BH}}$ at intermediate masses.
  Since the agreement with data at the high-mass end is satisfactory, this result in a different slope of the $\rm{M_{BH}-M_{Bulge}}$ relation with respect to observations.

It is worth stressing that the overall shape of the $\rm{M_{BH}-M_{Bulge}}$ relation depends on the whole accretion history of SMBHs.
  In particular, in \hen, the main mechanism responsible for both SMBH and bulge growth are galaxy mergers \citep{Kauffmann2000}, i.e. the so-called QSO-mode of SMBH accretion.
  \cite{Marulli2008} already showed the limitation of the \cite{Kauffmann2000} model in reproducing the redshift evolution of the bright QSO population.
  In this scenario, the radio-mode feedback contributes marginally to the shape of the relation.
  We do not expect the tension between \hen and observed $\rm{M_{BH}-M_{Bulge}}$ relation to affect our conclusions much, given the simple scaling of radio-mode efficiency with $\rm M_{BH}$ in \Equ{radiomode}.
  This is especially true at the high-mass end of the SMF, where the radio-mode prescription has been calibrated against.

Moreover, it is also worth noticing that the exact shape and scatter in the $\rm{M_{Bulge}-M_{BH}}$ relation have been revised by a number of works (\citealp{Graham2013}, \citealp{Fontanot2015} and \citealp{Shankar2016}), highlighting possible selection biases, that make the exact comparison between model predictions and data outside the aims of the present work.


\subsection{Impact on SFR Quenching}

Our analysis shows that, the lack of star formation activity in observed central galaxies correlates with observed stellar mass and the presence of a massive bulge component.
\Fig{passive_bubh} shows the behaviour of passive fractions for model central galaxies in $\rm M_{BH}-M_{Bulge}/M_{*}$ parameter space.
In the left panel we present the passive fraction for central galaxies in \hen.
Quenching in central galaxies is dominated by the presence of a massive black hole, independent of baryonic properties such as stellar mass.
As soon as the black hole mass reaches a certain threshold ($\rm{\log_{10}(M_{BH}/M_{\odot})\sim 6.0}$), quenching of star formation seems to onset.

In the right panel, the passive fraction for \sage central galaxies is presented.
In general, the passive fraction seem to be driven by the presence of massive black holes and a significant central bulge.
Star forming central galaxies in \sage have relative small black holes and small bulges.
It is only when both the black hole and bulge become gravitationally significant, the onset of quenching occurs.
However due to lack of big statistics in observed SMBH masses, \Fig{passive_bubh} can only be presented for SAMs and therefore, remains a prediction.


\section{Conclusion}
In this work, we have presented how star formation quenching depends on various galaxy properties and their environment for two SAMs and an observed sample. 
Theoretically, most massive galaxies reside in the deepest part of the gravitational potential well which corresponds to the centre of the dark matter halo.
In such cases, one should expect that these systems accrete most of the gas through cooling flows from the hot gas reservoir and hence go through a continuous star formation activity.
Massive galaxies also host  massive central black holes that inject energy in the system through radio mode AGN feedback.
In most cases, this energy will be enough to completely suppress the cooling flow resulting in star formation quenching.

The radio mode AGN feedback is an empirical prescription embedded in SAMs to deposit energy into the galaxy halo (see equations \Equ{radiomode} and \ref{equ:c16radio}).
\hen and \sage implement radio mode AGN feedback that offsets the gas cooling in massive central galaxies that live in dense environment, quenching them (see \Fig{den_hot_cooling}).
In order to study the impact of radio mode feedback, global properties such as halo mass and stellar mass are of great importance.
The known stellar - halo mass relation implies that massive galaxies reside in massive halos and observationally have low star formation activity \citep{conroy2009, behroozi2013, kravstov2018}.
But due to the strong correlation between the stellar mass and halo mass of central galaxies, it becomes difficult to decide whether it is the halo mass or stellar mass that dominates quenching of star formation.
We then use an observationally motivated parameter, the environmental density in a cylindrical aperture introduced in \cite{fossati2015}, to study star formation quenching.
Environmental density breaks the degeneracy between halo mass and stellar mass and can easily be implemented on both observed and model galaxies.
We use the same cylindrical aperture to select central galaxies from the models and observations; we also assume an adaptive aperture to select a pure sample of central galaxies for the analysis.

The evidence that massive observed central galaxies are quenched suggests that presence of mechanisms the suppress gas cooling or eject gas to cease star formation.
Both SAMs do not ideally match the observed passive fractions which correlate with the stellar mass and bulge mass.
Central passive fractions in \hen correlate with model halo and bulge mass whereas for \sage, passive fractions correlate with SMBH and bulge mass.
Both SAMs predict star forming massive isolated galaxies which is in contrast with observations.
Furthermore, \hen has problems with black hole growth in an isolated environment; massive field galaxies seem to have SMBH two orders of magnitudes smaller than \sage.
Such a discrepancy is very evident in the intermediate mass range of the $\rm{M_{BH}-M_{Bulge}}$ relation.
In order to match the observation, we suspect that SMBHs in isolated massive centrals in \hen should grow faster than their bulges.

Meanwhile, the suppression of cooling flows in \sage via radio mode feedback is less effective than the \hen framework.
In \sage, massive central galaxies that live in dense environment are forming stars at higher rate compared to \hen.
This can be a result of higher merger rates that bring in a large amount of cold gas that is available for star formation.
Furthermore, the high star formation rate is attributed to a higher star formation efficiency parameter in the Kennicutt-Schmidt relation \citep{kennicutt98}.

While both SAMs have their strengths and weaknesses, the simple description of AGN feedback in \hen seems to be performing as good as with respect to a more complex treatment in \sage.
However, a better treatment for SMBH growth is required for \hen that can allow for a better agreements with observed $\rm{M_{BH}-M_{Bulge}}$ relation.

\section*{Acknowledgements}

NA is grateful to the Queen's University, Canada for support through various scholarships and grants.
DJW acknowledges the support of the Deutsche Forschungsgemeinschaft via Projects WI 3871/1-1, and WI 3871/1-2.
MF has received funding from the European Research Council (ERC) under the European Union's Horizon 2020 research and innovation programme (grant agreement No 757535).
NA thanks Connor Stone and Mark Richardson for illuminating discussions.




\bibliographystyle{mnras}
\bibliography{reference} 




\appendix

\section{Selection of Central Galaxies}
\label{sec:central_select}

\begin{figure*}
\begin{multicols}{2}
    \includegraphics[width=\linewidth]{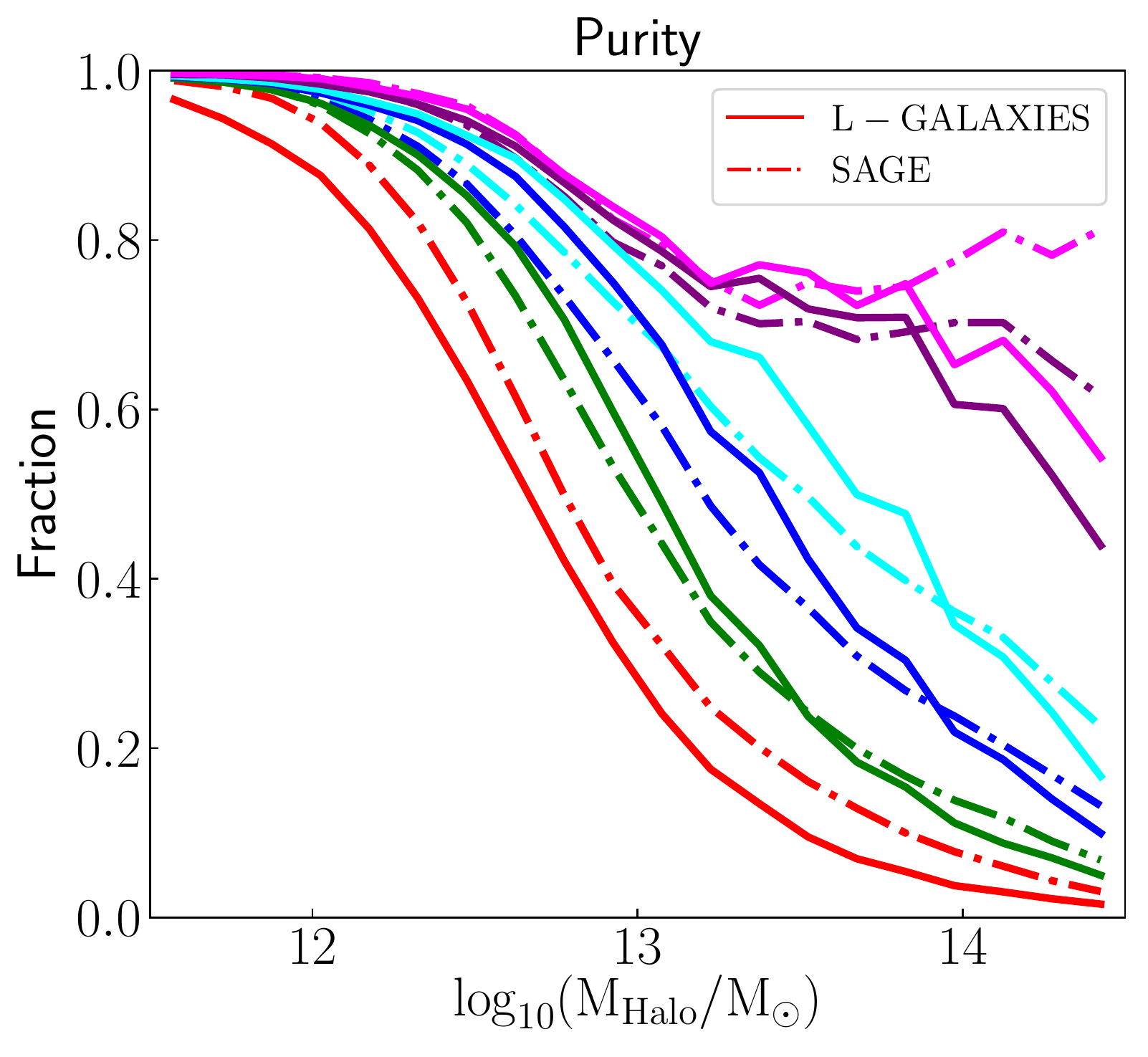}\par 
    \includegraphics[width=\linewidth]{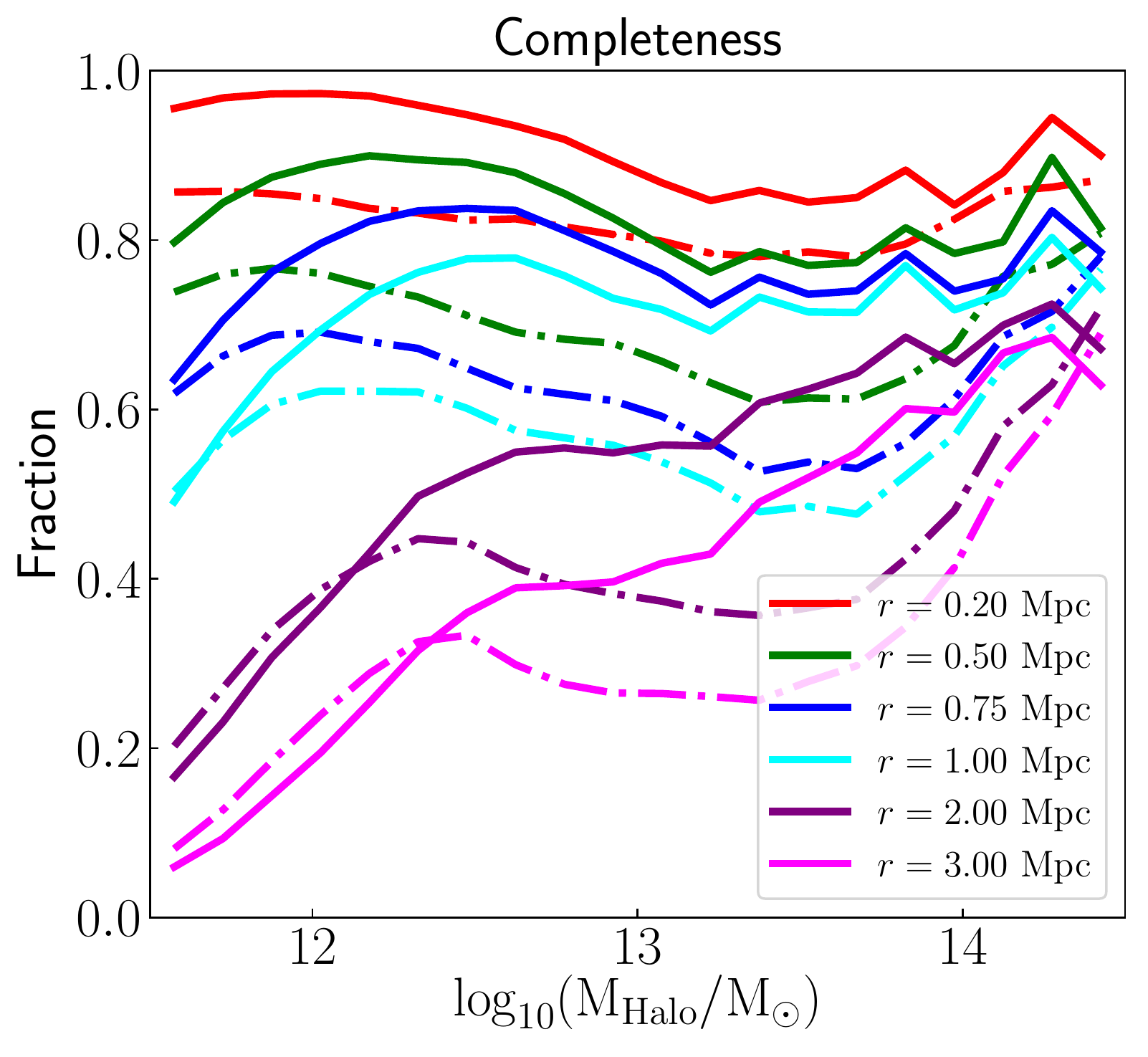}\par 
\end{multicols}
\caption{Purity and completeness as a function of halo mass for central galaxies (Mass Rank$~= 1$). The different colours show various aperture radius where the velocity depth has been fixed to $1000~\rm{km~s}^{-1}$. Solid lines show  {\scriptsize L-GALAXIES} and dashed-dotted lines show central galaxies from \sage.}
\label{fig:simPC}
\end{figure*}

Central galaxies are expected to be gravitationally dominant in the environment that they live in.
Star formation quenching in central galaxies is mainly caused by the energy outflows due the presence of an AGN \citep{silk2012}.
Therefore, an accurate selection of central galaxies is critical.
Any contamination in the central galaxy selection can affect the trends seen in passive fractions and leads to inaccurate conclusions about the impact of AGN heating. 

Before starting the discussion about the selection of central galaxies, we define two metrics that characterize a statistical sample: purity (P) and completeness (C).
The purity is defined as the ratio of the number of correctly identified centrals and the total number of identified centrals;  the completeness is the ratio of identified centrals and the total number of central galaxies.
Maximizing both quantities is desired, however a trade-off between the two needs to be found for selecting central galaxies \citep{fossati2015}.

The galaxy formation models provided the information about a galaxy being central or satellite.
For observed systems, we can use observed parameters to investigate its gravitational dominance.
To select central galaxies, we place a cylindrical aperture with radius in physics and the height in velocity space.
Each galaxy in this cylinder is assigned a rank based on its stellar mass.
If a galaxy achieves a mass rank of 1, it is classified as a central galaxy.
The radius of the cylinder is either fixed or adaptive aperture, as defined \Equ{adaptive} where $\rm r_{max}$ is the maximum radius the adaptive aperture can have, $\rm \alpha$ and $\rm \beta$ are parameters the relate the virial radius of the halo to the stellar mass.
The parameter $ n$ is defined as the isolation criterion and plays the role of preventing small apertures that might lead to a decrease in purity.
 
\begin{equation}
\rm{r(n, r_{max}, v_{depth}) = min(r_{max}, n\cdot 10^{\alpha \log M_*+\beta})~[Mpc]}
\label{equ:adaptive}
\end{equation}

We start with the discussion of selecting central galaxies while having the radius and velocity of the projected cylinder fixed.
\Fig{simPC} describes the purity (left) and completeness (right) for selecting central galaxies ($\rm {Mass Rank} = 1$) as a function of halo mass for the two simulations (solid: \hen and dashed-dotted: \sage).
The colours show the various radii of the projected aperture while the velocity depth is kept constant at $1000~\rm{km~s}^{-1}$.

\begin{figure*}
\begin{multicols}{2}
    \includegraphics[width=\linewidth]{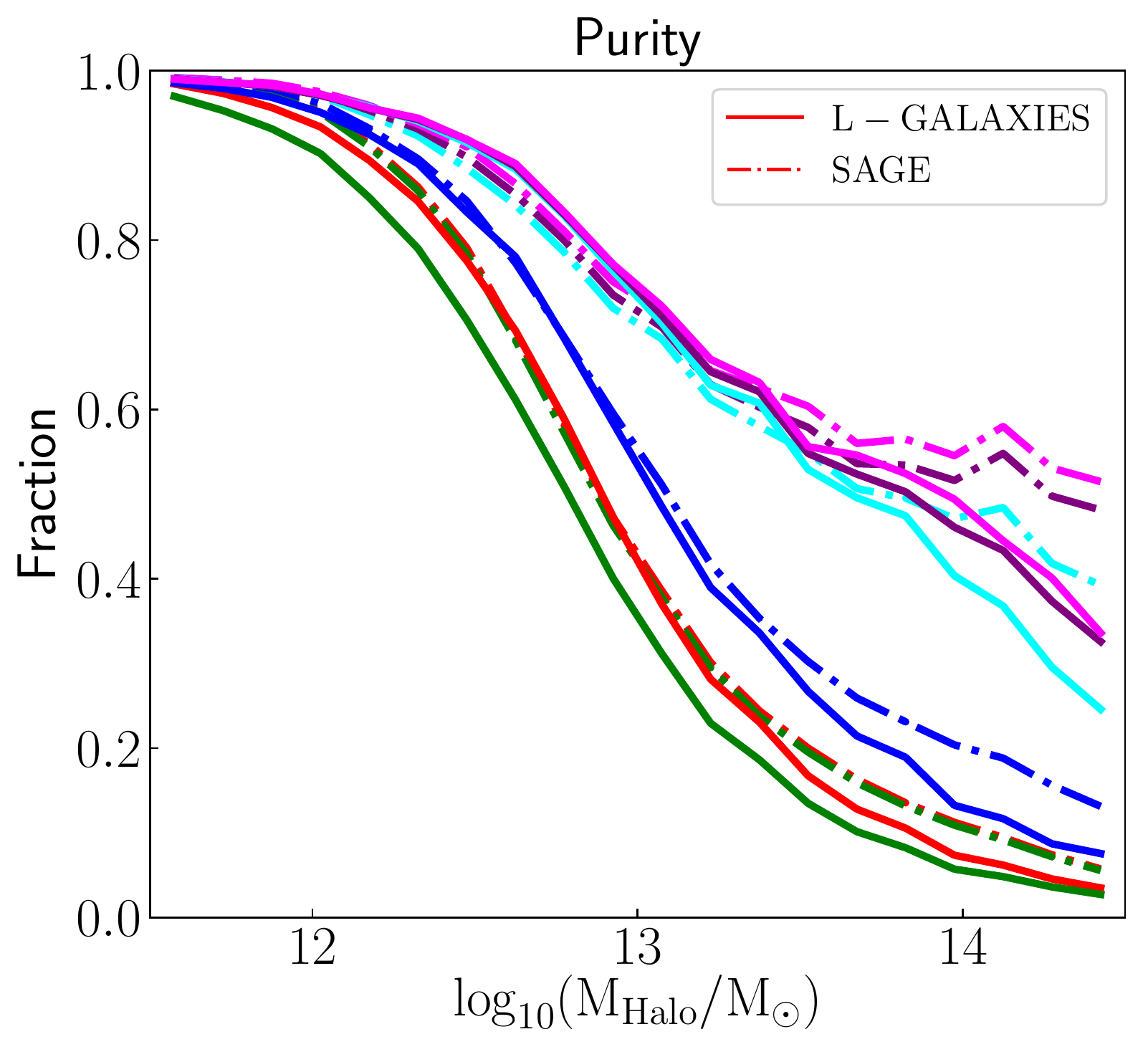}\par
    \includegraphics[width=\linewidth]{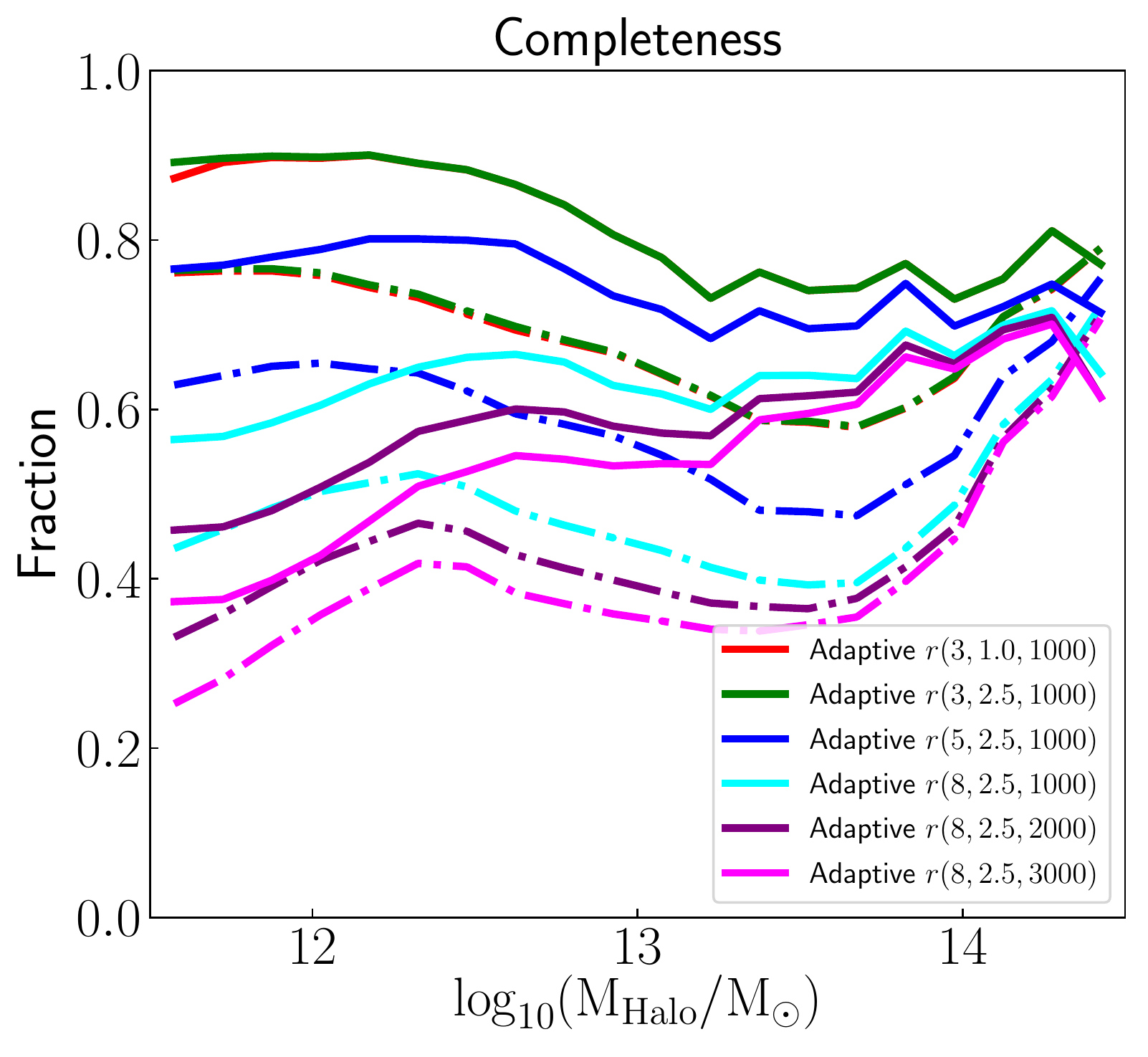}\par
\end{multicols}
\caption{Purity and completeness as a function of halo mass for central galaxies (Mass Rank$~= 1$). The different colours show various adaptive\
 aperture represented using \Equ{adaptive}. Solid lines show \hen and dashed-dotted lines show central galaxies from \sage.}
\label{fig:adaptivePC}
\end{figure*}

A small aperture results in most galaxies being classified as central galaxy which, of course, is incorrect.
The smaller aperture is unreliable at high halo masses where central galaxies should live in dense environment, i.e. high number of satellite galaxies.
With bigger apertures the purity at high halo mass increases as satellites are correctly identified. Completeness as a function of halo mass, for small radius, decreases on slightly.
As we increase the aperture size, the completeness at low halo masses decreases faster than at high halo masses.
The completeness of our algorithm never reaches $1.0$ because the most massive galaxies in halos might not be the central galaxy \citep{skibba2011}.
In general as the size of the aperture increases, the purity increases and completeness decreases for both simulations.
Our results are consistent with that of \cite{fossati2015} which used a galaxy formation model from \cite{guo2011}.
With fixed aperture, finding a balance between reasonable purity and completeness seems to be a difficult task.

In order to find an optimal balance between the purity and completeness, we make use of the adaptive aperture.
The adaptive aperture exploits the strong correlation between the stellar mass and the size of the dark matter halo for central galaxies.
\Fig{adaptivePC} depicts the various configurations of the adaptive apertures that are tested.
Similar trends are seen for the purity and completeness in the various adaptive apertures as compared to the fixed aperture.
\Fig{adaptivePC} tells us that an increase in the value of $n$ improves the purity of the selection at high halo masses.
On the other hand, a sharp decrease in completeness is seen for the central sample at low halo masses.
Independent of $n$, the velocity depth does not show any impact on the purity and completeness of the sample.
While we put more emphasis on the purity of the sample, a certain balance between purity and completeness is desired.
Therefore for both our simulations, we use the adaptive aperture with $\rm{n=8}$, $\rm{r_{max}=2.5~Mpc}$ and $\rm{v_{depth}=2000km~s^{-1}}$ to select central galaxies for the rest of the study.

As the input parameters for selecting central galaxies can be easily measured observationally, this technique can be implemented on both model and observed galaxies.
The uniform comparison of model and observed galaxies is a great advantage of this technique.
However, not all central galaxies are the most massive which results in some satellites being defined as centrals.
For example, in small galaxy groups where the central and satellite galaxy have similar stellar masses, our definition of central galaxy might not be accurate \citep{fossati17}.
For further details about the shortcomings of our techniques, the reader is referred to \cite{fossati2015}.



\bsp	
\label{lastpage}
\end{document}